\documentclass[11pt, a4paper]{article}

\usepackage[margin=1in]{geometry}
\usepackage{amsmath, amssymb, amsfonts, amsthm}
\usepackage{graphicx}
\usepackage{booktabs}
\usepackage{setspace}
\usepackage[natbibapa]{apacite}
\usepackage[hidelinks]{hyperref}
\usepackage{microtype}

\shortcites{%
  Chen2026, albert1984, baker2004, ban2001, bandalos2018, belzak2026,
  carroll2006, chalmers2012, chen2017, chenwang2016, cureton1966, deayala2009,
  eggen2011, firth1993, frazier2025, gelman2008, gignac2025, glas1988, guo2011,
  guo2020, he2026, heetal2017, heinze2002, henrysson1963, jacob2017, jewsbury2020,
  jewsbury2023, jewsbury2024, jewsbury2025, jewsburyetal2025, jewsburyjia2024, jonesjin1994,
  kennepagui2017, kim2006, kolen2014, kosmidis2020, liu2009, liu2025, lord1980, lord1983,
  magis2017, marsman2016, meng1994, mislevy1991, mislevy1996, murphytopel1985,
  naismith2025, neweymcfadden1994, nydick2026fixed, plummer2015, rubin1987, serfling1980,
  sharpnack2026, stefanskicarroll1985, stocking1988, swaminathan1986, vondavier2009,
  vondavier2024, wainermislevy1990, weiss2011, zijlmans2019%
}

\newcommand{\op}{\mathrm{op}}
\newcommand{\pilot}{\mathrm{pilot}}
\newcommand{\post}{\mathrm{post}}
\newcommand{\sub}{\mathrm{sub}}
\newcommand{\true}{\mathrm{true}}
\newcommand{\corr}{\mathrm{corr}}
\newcommand{\uncorr}{\mathrm{uncorr}}

\newcommand{\PELC}{\mathrm{PELC}}
\newcommand{\MLC}{\mathrm{MLC}}

\newtheorem{proposition}{Proposition}
\newtheorem{assumption}{Approximation}
\newtheoremstyle{condition}{\topsep}{\topsep}{\itshape}{}{\bfseries}{.}{.5em}{}
\theoremstyle{condition}
\newtheorem{condition}{Condition}

\title{Analytically Corrected Bayesian Modularization for Local Item Calibration\textsuperscript{*}}
\author{Paul A. Jewsbury\textsuperscript{\dag} and Steven W. Nydick\textsuperscript{\ddag} \\ Duolingo}
\date{}
 
\begin{document}
 
\maketitle
{\renewcommand{\thefootnote}{\fnsymbol{footnote}}%
\footnotetext[1]{We thank J.R. Lockwood and Alina von Davier for comments on an earlier draft of this article.}%
\footnotetext[2]{\textbf{Corresponding author.} Correspondence concerning this article should be addressed to Paul A. Jewsbury, Duolingo, Inc., 5900 Penn Avenue, Pittsburgh, PA 15206, United States. Email: \texttt{paul.jewsbury@duolingo.com}. ORCID: \href{https://orcid.org/0000-0001-5571-4623}{0000-0001-5571-4623}.}%
\footnotetext[3]{ORCID: \href{https://orcid.org/0000-0002-2908-1188}{0000-0002-2908-1188}.}}%
\setcounter{footnote}{0}

\begin{abstract}
The continuous calibration of pilot items embedded in operational assessments is challenging when pilot samples are small and adaptively routed. We formalize a Bayesian modularization framework for local item calibration that blocks feedback from pilot responses to the operational latent scale: Plausible Values are drawn from the operational posterior, and each pilot item is calibrated by its own local logistic regression. This construction is computationally scalable, protects operational trait estimates from malfunctioning pilot items, and admits Firth's penalized likelihood for sparse routed samples. Because treating imputed traits as fixed predictors induces attenuation, we derive closed-form disattenuation mappings that recover the generating item parameters under normal-ogive, posterior-normality, homoscedasticity, and joint-normality approximations. The resulting Modular Local Calibration (MLC) estimator reaches the target of marginal-likelihood calibration without per-item numerical integration. A multivariate delta-method covariance combined with Rubin's-rules pooling propagates operational item-parameter uncertainty into the focal item standard errors. Monte Carlo simulations for the unidimensional 2PL show that corrected MLC substantially reduces attenuation bias and yields nominal-to-conservative interval coverage in the studied conditions, including under restricted-range MAR routing.
\end{abstract}


\section{Introduction}

Operational testing programs must continuously calibrate new assessment items to ensure validity, test security, and scale stability \citep{Chen2026, deayala2009, he2026}. To gather response data free from motivational artifacts, programs routinely embed uncalibrated pilot items into live operational test administrations \citep{belzak2026, vondavier2024}. A primary purpose of this embedded data collection is to evaluate the psychometric properties of these new items to screen out defective content. However, in modern adaptive testing designs, the routing algorithms that assign pilot items rely on the examinee's observed operational responses \citep{naismith2025}. Consequently, calibrating pilot items in isolation makes the missingness mechanism nonignorable for any model that conditions only on pilot responses, which can induce estimation bias \citep{eggen2011, glas1988, jewsbury2020, jewsbury2025, mislevy1996}. 

To satisfy the Missing at Random (MAR) assumption \citep{rubin1976} and yield asymptotically unbiased parameter estimates, the calibration model must condition on the operational responses \citep{jewsbury2024, mislevy1996}. Under standard full-information (FI) item response theory (IRT) frameworks, whether marginal maximum likelihood (MML) or fully Bayesian Markov chain Monte Carlo (MCMC), this requires joint calibration of the sparse pilot responses with the large operational bank. However, as operational pools can scale to tens of thousands of items (e.g., the Duolingo English Test; \citealp{naismith2025}), joint calibration is computationally burdensome at scale, whether through large-scale optimization or high-dimensional posterior sampling.

\subsection{The Limitations of Proxy and Point-Estimate Calibration}

Historically, practitioners have bypassed the computational burden of FI IRT when estimating the measurement parameters of a single item by replacing the unobserved latent trait ($\theta$) with a manifest proxy, such as the operational sum score ($X$) or an interim point estimate \citep{gignac2025, zijlmans2019}. This enables Classical Test Theory (CTT) heuristics to flag under-performing items, particularly the item-total biserial correlation, which can be mapped to IRT discrimination parameters via normal-ogive approximations \citep{lord1980}. 

Substituting a point estimate for the latent trait introduces three statistical distortions. First, under logistic IRT models, the true relationship between the manifest score and $\theta$ is nonlinear, violating the assumptions of Pearson correlations \citep{guo2020}. Second, adaptive routing's restriction of trait variance in the targeted subset results in artificially lower discrimination magnitudes \citep{magis2017, weiss2011}. Third, and central to this paper, manifest scores and interim point estimates contain inherent measurement error: treating these proxies as error-free covariates ignores the posterior variance of the trait, producing classical attenuation bias and anti-conservative standard errors that compromise subsequent inference \citep{bandalos2018, guo2011}. 

\subsection{Online Calibration in Computerized Adaptive Testing}

Calibrating new items from ability information supplied by already-calibrated operational items has been studied extensively in computerized adaptive testing (CAT), where it is termed \textit{online calibration} \citep{stocking1988}. In a CAT, each examinee's ability is estimated from the operational items already administered, and uncalibrated pretest items are seeded into the administration to be calibrated against those estimates. For this task, \citet{stocking1988} introduced \textit{Method A}, also called Fixed Ability Parameter Calibration (FAPC; \citealp{ban2001}). Method A treats the operational maximum likelihood estimate (MLE) of ability as a fixed, error-free predictor and calibrates each new item by an independent logistic regression.

Method A is the prototypical instance of \textit{local item calibration}: a procedure that calibrates each new item separately, in its own low-dimensional parameter space, against ability estimates supplied by the already-calibrated operational items, rather than through a single concurrent estimation that couples all item and person parameters. The qualifier \textit{local} refers to this per-item scope (one item calibrated at a time), and it is the property that has made Method A among the simplest and most widely used online-calibration procedures.

Treating the ability estimate as error-free, however, ignores its measurement error and attenuates the estimated discrimination toward zero, the classical effect of conditioning on a noisy predictor. Correcting this attenuation, rather than replacing Method A with an integration-based alternative such as those discussed below, is a more recent development. Two corrections operate on the ability estimate before it enters the calibration. \citet{chenwang2016} augment Method A with the full functional maximum likelihood estimator (FFMLE) of \citet{jonesjin1994} and \citet{stefanskicarroll1985}, which models the operational ability estimate as carrying measurement error. \citet{heetal2017} instead iterate Lord's \citeyearpar{lord1983} bias correction on the operational MLE before it is used as the predictor. The two target different properties of the ability estimate. Method A's attenuation is governed by the variance of the operational estimate rather than its bias, so reducing the bias leaves the attenuation largely in place, whereas the functional measurement-error correction addresses the variance directly. \citet{chenwang2016} also administer pilot items by random selection and leave adaptive routing of pilot items to future work. 

\citet{wainermislevy1990} introduced OEM (one expectation--maximization cycle), which estimates new-item parameters by marginal maximum likelihood with the trait posterior conditioned only on operational responses. Like Method A, OEM is a local item calibration, constructing the trait estimate without new-item responses, but it integrates over the full operational posterior rather than treating a point estimate as fixed. \citet{ban2001} introduced MEM (multiple EM cycles), which iterates with the trait posterior conditioning on both response blocks from the second iteration. Comparative work, including the multidimensional and Bayesian variants of \citet{chen2017}, finds the relative ordering of the performance of OEM, MEM, and Method A depends on sample size, dimensionality, and inter-dimensional correlation.

These methods differ in the extent to which new-item responses enter the trait estimate used to calibrate them. Method A and OEM construct the trait estimate from operational responses only, providing full structural protection from tautological contamination. FFMLE-M-Method A is intermediate: the initial trait estimate uses operational responses only, and the per-pilot Newton update subsequently incorporates the focal pilot's responses into the corrected ability used in the score equation. MEM and joint FI MML restore the new-item-to-trait feedback channel iteratively or simultaneously, forfeiting modular contamination protection entirely. The Bayesian modularization framework introduced next formalizes this distinction and unifies the existing methods with the procedure developed here.

\subsection{Bayesian Modularization and Local Calibration}

We adopt a formal Bayesian modularization perspective \citep{frazier2025, jacob2017, liu2009, liu2025}. Bayesian modularization severs the feedback between components of a joint model, so that a designated ``trustworthy'' module can be estimated without contamination from a ``suspect'' one. Here, the trustworthy module comprises the operational items and person parameters, and the suspect module the uncalibrated pilot items; we refer to this instantiation as \emph{local-calibration modularization}. With the person parameters estimated conditional on the operational data alone, the high-dimensional concurrent calibration decomposes into independent, parallelizable per-pilot logistic regressions. Analogous modularization logic underlies stepwise estimation of latent variable models \citep{vermunt2025}, most closely the Bayesian measurement and uncertainty preserving parametric (MUPPET) approach \citep{levy2023, levy2025, levymcneish2025}, which cuts structural-to-measurement feedback while propagating measurement uncertainty forward. The same logic has been applied to IRT linking in fixed parameter chain anchoring (FPCA; \citealp{nydick2026fixed}), which fixes anchor-item parameters at samples drawn from their calibration posterior and pools across the resulting chains to propagate anchor uncertainty into the calibration of new items.

In this paper, local-calibration modularization formalizes two common psychometric practices designed to prevent contamination. First, in standard IRT equating and pretesting, fixed item parameter calibration holds the parameters of well-estimated operational items constant, preventing the latent scale from being degraded by joint estimation of uncalibrated pilot items \citep{kim2006, kolen2014}. Second, in classical item analysis, the focal item is excluded from the composite proficiency estimate when calculating corrected item-total correlations to prevent spurious inflation and tautological contamination \citep{cureton1966, henrysson1963}, the classical-test-theory precedent for the same focal-item exclusion that the modular framework imposes via posterior cuts. Bayesian modularization formalizes this protective decoupling within a unified model and extends it from a heuristic adjustment in CTT to a coherent justification in IRT.

Beyond computational speed and contamination protection, collapsing calibration to localized two-dimensional parameter spaces provides an additional advantage for sparse data. Sparse pilot samples are prone to complete or quasi-complete separation, in which trait level perfectly or almost perfectly divides correct from incorrect responses and no finite maximum likelihood estimate exists. In FI joint estimation, the standard remedies are ad hoc boundary constraints or weakly informative subjective priors \citep{gelman2008, swaminathan1986}. The proposed localized method instead reduces each pilot calibration to a standard two-dimensional logistic regression. For this regression, Firth's penalized likelihood \citep{firth1993} operationalizes the objective Jeffreys prior and yields finite estimates without subjective tuning \citep{heinze2002}. It is directly available in widely used software \citep{kosmidis2020}.

\subsection{The Current Study}

Local-calibration modularization achieves operational scalability, contamination protection, and sparse-data penalization, but the local regressions condition on a fallible proxy for the latent trait as though it were error-free. That proxy is constructed from the operational module alone, the same omission of pilot responses that prevents tautological contamination. The construction is uncongenial in the sense of \citet{meng1994}, and the unmodeled posterior variance of the trait attenuates the item-parameter estimates. Correcting this attenuation while preserving the modular structure is a key problem this paper addresses.

This paper develops local-calibration modularization into an online-calibration procedure that retains these advantages while analytically removing the attenuation the framework's fallible trait proxy induces. Three contributions, developed in Section~2 and evaluated in Section~3, support this claim. First, we formalize two estimators within the framework: Modular Local Calibration (MLC), defined by multiple imputation over Plausible Values drawn from the operational posterior, and its point-estimate counterpart, Point Estimate Local Calibration (PELC). Because the imputation conditions on operational responses alone, the framework satisfies MAR for adaptively routed pilots, blocks tautological contamination from malfunctioning items, decomposes the concurrent calibration into independent per-pilot logistic regressions that scale to operational banks of tens of thousands of items, and supports Firth's penalized likelihood natively for separation control under sparse routing \citep{firth1993}. Second, we derive closed-form disattenuation mappings that recover the generating item parameters at the population level under stated approximations, evaluated on the realized routed subset so as to absorb the trait-location and range-restriction effects adaptive routing induces: effects prior online-calibration corrections do not address. We prove the corrected estimator recovers the generating parameters (Proposition~\ref{prop:inverse}) and so shares the modular marginal-likelihood target OEM \citep{wainermislevy1990} reaches by per-item numerical integration. Its distinguishing feature is that it reaches that target as an algebraic transform of the logistic regressions an operational scoring pipeline already computes, so the trait-side cost amortizes across the item bank rather than recurring as a separate integration for each pilot item. Third, we develop a standard-error construction pairing a multivariate Jacobian delta-method scaling with Rubin's-rules pooling that, by sampling operational item parameters from their calibration posterior, propagates operational-bank estimation uncertainty into the new-item standard errors. Section~3 establishes these properties by Monte Carlo for the unidimensional 2PL: approximately unbiased recovery of $(a_j, d_j)$ and nominal-to-conservative interval coverage across data-sparsity and routing conditions, including adaptive routing.

\section{Bayesian Modularization of Item Calibration}

To resolve the computational and statistical bottlenecks of joint FI estimation, we formalize a modular framework that decouples pilot-item estimation from the operational latent scale. 

Operational programs calibrate the item bank before pilot data are available and then hold the operational item parameters fixed for scoring. Two disjoint examinee samples are involved. A \emph{calibration sample} of $N_{\op}$ examinees, each administered $J_{\op}$ operational items, identifies the operational item parameters $\boldsymbol{\eta}_{\op}$ (each item carrying its own parameter count and response model). We write $\mathbf{Y}_{\op}$ for its $N_{\op}\times J_{\op}$ response matrix, which enters the analysis only through the posterior $p(\boldsymbol{\eta}_{\op}\mid\mathbf{Y}_{\op})$. A disjoint \emph{pilot-administration sample} $\mathcal{P}$ is scored on the operational items at these fixed estimates and routed to the pilot items, its operational responses never re-entering the calibration of $\boldsymbol{\eta}_{\op}$. Writing $\mathbf{y}_{\op,i}$ for the $1\times J_{\op}$ operational response vector of examinee $i\in\mathcal{P}$, each pilot item $j$ is administered only to a routed subset $\mathcal{I}_j=\{i\in\mathcal{P}: Y_{ij}\text{ observed}\}$, with $N_j=|\mathcal{I}_j|$ and parameters $\boldsymbol{\eta}_j$. Full-information calibration would update the traits, $\boldsymbol{\eta}_{\op}$, and the pilot parameters jointly from all responses, re-estimating the operational scale whenever pilots are added. The modular framework instead conditions on the fixed operational module, calibrating each pilot without recalibrating the bank. With the feedback from pilot responses to the latent trait severed, the focal pilot's parameters are estimated by maximizing the modular marginal likelihood
\begin{equation}
L_{\mathrm{mod}}(\boldsymbol{\eta}_j) \;=\; \prod_{i \in \mathcal{I}_j} \int p(y_{ij} \mid \theta_i, \boldsymbol{\eta}_j)\, p(\theta_i \mid \mathbf{y}_{\op,i}, \boldsymbol{\eta}_{\op})\, d\theta_i,
\label{eq:mod_likelihood}
\end{equation}
which identifies the generating parameters $(a_j, d_j)$ under correct specification and is the population target reached directly by OEM \citep{wainermislevy1990} via per-item numerical integration; the framework developed below reaches the same target through ordinary local logistic regressions on Plausible Values followed by a closed-form algebraic correction.

This modular framework yields two practical advantages. First, it eliminates tautological contamination: poorly functioning pilot items cannot distort the proficiency estimates used to evaluate them. Second, because pilot routing depends only on examinees' operational responses, conditioning on the operational module supports a MAR formulation for pilot-item administration \citep{jewsbury2025, mislevy1996}.

\subsection{Imputation of the Trustworthy Module}

Under the modular framework, the posterior of the pilot-item parameters has no closed form and is not straightforward to sample directly. Modularization is therefore operationalized via posterior imputation \citep{plummer2015}: the latent proficiency vector is integrated out by drawing independent Plausible Values (PVs) from the posterior of $\boldsymbol{\theta}$.

Plausible-values methods have a substantial literature in large-scale survey assessment \citep{mislevy1991, jewsbury2023, jewsburyetal2025, marsman2016, vondavier2009}, where the inferential target is typically a feature of the latent distribution with IRT parameters as fixed nuisance. The present application reverses this: the latent trait is the nuisance and the new pilot-item parameters are the targets. The plausible-values literature explicitly cautions against secondary analyses whose targets fall outside the imputation model's conditioning set \citep{jewsburyjia2024, marsman2016, vondavier2009}; calibrating an item on Plausible Values that exclude that item's responses is such an analysis, and the closed-form correction of Section~2.4 is what converts this cautioned-against use into a consistent one. Under this reversal, congeniality \citep{meng1994} requires inclusion of the focal pilot's responses in the imputation model that generates the trait estimates against which the same item is calibrated: precisely the contamination the modular framework is designed to prevent. The deliberate uncongeniality is therefore a desirable property, with the analytic correction of Section~2.4 the price paid for it. To our knowledge, this multiple-imputation-with-PVs construction is novel for new-item calibration.

To propagate operational-bank estimation uncertainty into the pilot SEs, the imputation incorporates the operational-parameter posterior. For each imputation $s\in\{1,\dots,S\}$, operational item parameters $\boldsymbol{\eta}_{\op}^{(s)}$ are drawn from their calibration posterior $p(\boldsymbol{\eta}_{\op}\mid\mathbf{Y}_{\op})$, and each scored examinee's proficiency is sampled conditional on that draw as well as the examinee's own operational responses,
\begin{equation}
    \theta_i^{(s)} \sim p(\theta_i \mid \mathbf{y}_{\op,i}, \boldsymbol{\eta}_{\op}^{(s)}),
    \qquad i \in \mathcal{P}.
\end{equation}
Drawing $\boldsymbol{\eta}_{\op}^{(s)}$ from the calibration posterior rather than fixing it at a point estimate is what propagates operational-bank estimation uncertainty into the pilot standard errors.

All summary statistics in the derivations that follow are local to the routed subset $\mathcal{I}_j$ defined above. Moreover, unhatted symbols throughout the remainder of this paper denote population quantities and hatted symbols denote empirical plug-ins from the imputed Plausible Values. The empirical latent mean and variance for subset $j$, $\hat\mu_{\sub,j}$ and $\hat\sigma^2_{\sub,j}$, converge in probability under the asymptotics of Section~2.5 to $\mu_{\sub,j} = E[\theta_i \mid i \in \mathcal{I}_j]$ and $\sigma^2_{\sub,j} = \mathrm{Var}(\theta_i \mid i \in \mathcal{I}_j)$.

The expected posterior variance on the routed subset, $\sigma^2_{\post,j} = E[\mathrm{Var}(\theta_i \mid \mathbf{y}_{\op,i}, \mathbf{Y}_{\op}) \mid i \in \mathcal{I}_j]$, is the population quantity that drives the attenuation in Sections~2.3.2--2.3.3; its subset-level analogue is the average $N_j^{-1} \sum_{i \in \mathcal{I}_j} \mathrm{Var}(\theta_i \mid \mathbf{y}_{\op,i}, \mathbf{Y}_{\op})$. By the law of total variance,
\begin{equation}
\mathrm{Var}(\theta_i \mid \mathbf{y}_{\op,i}, \mathbf{Y}_{\op})
\;=\;
E_{\boldsymbol{\eta}_{\op} \mid \mathbf{Y}_{\op}}\!\left[\mathrm{Var}(\theta_i \mid \mathbf{y}_{\op,i}, \boldsymbol{\eta}_{\op})\right]
\;+\;
\mathrm{Var}_{\boldsymbol{\eta}_{\op} \mid \mathbf{Y}_{\op}}\!\left[E(\theta_i \mid \mathbf{y}_{\op,i}, \boldsymbol{\eta}_{\op})\right],
\label{eq:totalvar_decomp}
\end{equation}
a partition of the posterior variance over draws of the operational item-parameter vector $\boldsymbol{\eta}_{\op}$ from its calibration posterior. The within-operational-parameter component, $E_{\boldsymbol{\eta}_{\op} \mid \mathbf{Y}_{\op}}[\mathrm{Var}(\theta_i \mid \mathbf{y}_{\op,i}, \boldsymbol{\eta}_{\op})]$, is the trait variance that remains when $\boldsymbol{\eta}_{\op}$ is held fixed at a single draw: the operational parameters are then known, no item-parameter uncertainty contributes, and the term is governed by the examinee's own operational test length $J_{\op}$. The between-operational-parameter component, $\mathrm{Var}_{\boldsymbol{\eta}_{\op} \mid \mathbf{Y}_{\op}}[E(\theta_i \mid \mathbf{y}_{\op,i}, \boldsymbol{\eta}_{\op})]$, is the additional variance contributed as the conditional posterior mean $E(\theta_i \mid \mathbf{y}_{\op,i}, \boldsymbol{\eta}_{\op})$ varies with the draw of $\boldsymbol{\eta}_{\op}$; it is the channel through which item-parameter uncertainty enters, governed by the calibration sample size $N_{\op}$ through the spread of $p(\boldsymbol{\eta}_{\op} \mid \mathbf{Y}_{\op})$. The empirical analogue $\hat\sigma^2_{\post,j}$ is given in Section~3.2.

\subsection{Local Calibration Estimators}

With the latent trait proxies treated as fixed predictors, the estimation of the suspect module reduces to an independent \textit{local calibration} for each pilot item $j$, evaluated over its routed subset $\mathcal{I}_j$. While this modular framework generalizes to any parametric latent variable model, we operationalize it for the two-parameter logistic (2PL) item response model via local logistic regression. The 2PL item response probability is commonly conceptualized in terms of discrimination ($a_j$) and difficulty ($b_j$), but is parameterized here via an intercept ($d_j = -a_j b_j$):
\begin{equation}
    P(Y_{ij} = 1 \mid \theta_i) = \mathrm{logit}^{-1}(a_j \theta_i + d_j) = \mathrm{logit}^{-1}\{a_j (\theta_i - b_j)\}.
\end{equation}
The local parameter vector is the $2 \times 1$ column vector $\boldsymbol{\eta}_j = [a_j, d_j]^\intercal$. We propose two classes of local calibration. Both fit ordinary local logistic regressions on a $\theta$-proxy and apply the closed-form analytic correction of Section~2.4 post hoc; they are referred to collectively as \emph{the corrected estimators} hereafter, in contrast to the uncorrected modular estimator (Method A, FFMLE-M-Method~A) and the per-pilot quadrature estimator (OEM).

\subsubsection{Point Estimate Local Calibration (PELC)}
The most computationally straightforward approach regresses pilot responses $Y_{ij} \in \{0, 1\}$ on examinee-specific expected a posteriori (EAP) point estimates ($\bar{\theta}_i$) via standard logistic regression:
\begin{equation}
    P(Y_{ij} = 1 \mid \bar{\theta}_i) = \mathrm{logit}^{-1}(a_j \bar{\theta}_i + d_j).
\end{equation}
While computationally efficient, this treats the EAP as an error-free covariate, ignoring the latent uncertainty.

\subsubsection{Modular Local Calibration via Posterior Integration (MLC)}

To account for latent uncertainty ignored by PELC without fitting the modular marginal likelihood \eqref{eq:mod_likelihood} directly (as OEM does via per-item numerical integration, or as a fully Bayesian errors-in-variables formulation does via MCMC), the MLC approach uses posterior imputation. Specifically, $S$ ordinary logistic regressions are fit, one per Plausible Value vector, with coefficients pooled across imputations and corrected via the closed-form mapping of Section~2.4 to recover the generating parameters. Let $\mathbf{y}_j$ denote the $N_j \times 1$ vector of observed pilot responses for item $j$, and let $\boldsymbol{\theta}_{\sub,j}^{(s)}$ denote the corresponding $N_j \times 1$ sub-vector of imputed Plausible Values for examinees in $\mathcal{I}_j$ under imputation $s$. The $S$ imputations yield the equal-weight mixture of local posteriors, $p(\boldsymbol{\eta}_j \mid \mathbf{y}_j) = S^{-1} \sum_{s=1}^{S} p(\boldsymbol{\eta}_j \mid \mathbf{y}_j, \boldsymbol{\theta}_{\sub,j}^{(s)})$, which approximates the pilot-parameter posterior of the modular framework. The notation suppresses the conditioning on the operational data ($\{\mathbf{y}_{\op,i}\}_{i \in \mathcal{I}_j}$ and $\mathbf{Y}_{\op}$) that enters through the Plausible Values.

While generating these $S$ models is procedurally identical to handling missing data, the pooling mechanisms represent distinct theoretical paradigms. The first two variants (MLC-RR and MLC-MN) aggregate analytic summary statistics: the local point estimates ($\hat{\boldsymbol{\eta}}_j^{(s)}$) and their asymptotic covariance matrices. The third (MLC-EC) bypasses point estimates entirely, aggregating stochastic random draws ($\boldsymbol{\eta}_j^{(s)}$) from the local posteriors. We evaluate these three pooling rules:

\begin{enumerate}
    \item \textbf{MLC-RR (Rubin's Rules):} Within the classical Multiple Imputation framework \citep{rubin1987}, the pooled point estimate is the arithmetic mean of the $S$ local estimates: $\bar{\boldsymbol{\eta}}_j = \frac{1}{S} \sum_{s=1}^S \hat{\boldsymbol{\eta}}_j^{(s)}$. The total variance combines the $2 \times 2$ within-imputation covariance ($\mathbf{W}_j$) and between-imputation covariance ($\mathbf{B}_j$) matrices with a finite-imputation penalty:
    \begin{align}
        \mathbf{W}_j &= \frac{1}{S} \sum_{s=1}^S \widehat{\text{Var}}(\hat{\boldsymbol{\eta}}_j^{(s)}), \\
        \mathbf{B}_j &= \frac{1}{S-1} \sum_{s=1}^S (\hat{\boldsymbol{\eta}}_j^{(s)} - \bar{\boldsymbol{\eta}}_j)(\hat{\boldsymbol{\eta}}_j^{(s)} - \bar{\boldsymbol{\eta}}_j)^\intercal, \\
        \mathbf{T}_j &= \mathbf{W}_j + \left(1 + \frac{1}{S}\right) \mathbf{B}_j.
    \end{align}
    
    \item \textbf{MLC-MN (Mixture of Normals):} The PV-induced mixture is approximated as an equal-weight Gaussian mixture: $p(\boldsymbol{\eta}_j \mid \mathbf{y}_j) \approx \frac{1}{S} \sum_{s=1}^S \mathcal{N}(\hat{\boldsymbol{\eta}}_j^{(s)}, \widehat{\text{Var}}(\hat{\boldsymbol{\eta}}_j^{(s)}))$. The pooled point estimate is identical to MLC-RR; the total variance follows from the law of total variance applied to the equal-weight mixture, $\mathbf{T}_j^{MN} = \mathbf{W}_j + \tfrac{S-1}{S}\mathbf{B}_j$, omitting the $1 + S^{-1}$ inflation of classical Rubin's rules.
        
    \item \textbf{MLC-EC (Empirical Cut):} The PV-induced mixture is evaluated by Monte Carlo, taking one stochastic parameter draw from the local posterior conditional on each imputed trait vector: $\boldsymbol{\eta}_j^{(s)} \sim p(\boldsymbol{\eta}_j \mid \mathbf{y}_j, \boldsymbol{\theta}_{\sub,j}^{(s)})$. With MCMC local fitting, this pools random draws rather than summary statistics and avoids any normality approximation; with maximum-likelihood local fitting, the draw is taken from the asymptotic normal approximation $\boldsymbol{\eta}_j^{(s)} \sim \mathcal{N}(\hat{\boldsymbol{\eta}}_j^{(s)}, \widehat{\text{Var}}(\hat{\boldsymbol{\eta}}_j^{(s)}))$. The final point estimate and $2 \times 2$ variance are the empirical mean and covariance of the $S$ stochastic draws.
\end{enumerate}

\subsubsection{Penalized Likelihood for Sparse Data}
Testing programs frequently calibrate pilot items using small, sparse samples (e.g., $N_j \le 200$), where standard MLE is susceptible to complete or quasi-complete separation \citep{albert1984}. Because modularization collapses the parameter space to two dimensions per item, both PELC and MLC accommodate Firth's penalized likelihood \citep{firth1993}, operationalizing Jeffreys' invariant prior $|\mathbf{I}(\boldsymbol{\eta}_j)|^{1/2}$:
\begin{equation}
    \ell^*(\boldsymbol{\eta}_j^{(s)}) = \ell(\boldsymbol{\eta}_j^{(s)}) + \frac{1}{2} \ln |\mathbf{I}(\boldsymbol{\eta}_j^{(s)})|.
\end{equation}
This is implemented in standard software (e.g., the \texttt{brglm2} R package; \citealp{kosmidis2020}). The attenuation derivations of Section~2.3 are stated for the unpenalized local MLE; because Firth's penalty is $O(1)$ relative to the $O(N_j)$ log-likelihood, it does not alter the population limits in Propositions~\ref{prop:pelc_atten} and~\ref{prop:mlc_atten} and serves only finite-sample bias reduction and separation control. The closed-form correction of Section~2.4 is therefore applied to the penalized estimates and asymptotically targets the same population limit as the unpenalized derivation.

\subsection{Theoretical Properties: Uncongeniality and EIV Attenuation}

To evaluate the structural bias and variance, we partition the expected Fisher information blocks. Let $\ell_{\pilot}$ denote the log-likelihood of the pilot responses $\mathbf{y}_j$, $\ell_{\op}$ the log-likelihood of the routed examinees' operational responses $\{\mathbf{y}_{\op,i}\}_{i \in \mathcal{I}_j}$ given $\boldsymbol{\eta}_{\op}$, and $\ell_{\op}^+ = \ell_{\op} + \log p(\boldsymbol{\theta})$ the corresponding operational log-posterior kernel for those examinees' traits.

Let $\mathbf{I}_{\eta}$ denote the $2 \times 2$ expected total information matrix of the pilot item parameters given $\boldsymbol{\theta}$:
\begin{equation}
    \mathbf{I}_{\eta} = -E\left[\frac{\partial^2 \ell_{\pilot}}{\partial \boldsymbol{\eta}_j \partial \boldsymbol{\eta}_j^\intercal}\right].
\end{equation}
Let $\mathbf{I}_{\theta, \op}^*$ denote the $N_j \times N_j$ effective information about the routed examinees' traits $\{\theta_i\}_{i \in \mathcal{I}_j}$ from the operational test and trait prior, formed by the Schur complement that partials out operational item-parameter uncertainty:
\begin{equation}
    \mathbf{I}_{\theta, \op}^* = -E\left[\frac{\partial^2 \ell_{\op}^+}{\partial \boldsymbol{\theta} \partial \boldsymbol{\theta}^\intercal}\right] - \mathbf{I}_{\theta \eta_{\op}} \mathbf{I}_{\eta_{\op}}^{-1} \mathbf{I}_{\eta_{\op} \theta}
\end{equation}
where $\mathbf{I}_{\eta_{\op}}$ is the expected information for the operational item parameters and $\mathbf{I}_{\theta \eta_{\op}} = \mathbf{I}_{\eta_{\op} \theta}^\intercal$ is the trait-by-operational-item-parameter cross-information. The standard-normal independent prior assumed throughout contributes an identity matrix to the trait block. The calibration sample size $N_{\op}$ enters $\mathbf{I}_{\theta, \op}^*$ only through the magnitude of $\mathbf{I}_{\eta_{\op}}$ in the partialling term. These information blocks reappear in the modularization-efficiency comparison of Supplement~S7.

By omitting the pilot response $Y_{ij}$ from the imputation model to prevent contamination, the framework is theoretically uncongenial \citep{meng1994}: the local likelihoods fail to account for posterior variance, which on the routed subset satisfies $\sigma^2_{\post,j} \approx \frac{1}{N_j} \sum_{i \in \mathcal{I}_j} [(\mathbf{I}_{\theta, \op}^*)^{-1}]_{ii}$ under joint asymptotics. This unmodeled variance alters expected parameter recovery, as derived below.

The congenial alternative, a fully Bayesian errors-in-variables suspect module \citep{carroll2006} that treats individual traits as unobserved parameters with operational-posterior priors and integrates them out via MCMC, is consistent without correction but incurs a per-pilot $N_j$-fold integration cost that does not amortize the way the Plausible-Values construction of Sections~2.1--2.2 does.

\subsubsection{Approximation Assumptions for Analytic Attenuation}

To evaluate structural bias analytically and derive closed-form invertible corrections, we adopt four approximations. The approximations are stated at the population level on the routed subset; finite-sample analogues replace population quantities with empirical plug-ins.

\begin{assumption}[Normal-Ogive]\label{ass:normal_ogive}
The logistic response function is approximated by a scaled cumulative normal, $\mathrm{logit}^{-1}(x) \approx \Phi(x/D)$, where $D = 1.702$ \citep{lord1980}.
\end{assumption}

\begin{assumption}[Posterior Normality]\label{ass:posterior_normal}
The individual operational posterior is approximately Gaussian: $\theta_i \mid \mathbf{y}_{\op,i} \stackrel{\cdot}{\sim} \mathcal{N}(\bar\theta_i, \mathrm{Var}(\theta_i \mid \mathbf{y}_{\op,i}))$, where $\bar\theta_i = E(\theta_i \mid \mathbf{y}_{\op,i})$ is the marginal posterior mean.
\end{assumption}

\begin{assumption}[Homoscedasticity on the Routed Subset]\label{ass:homosc}
The marginal posterior variance is approximately constant across examinees in the routed subset: $\mathrm{Var}(\theta_i \mid \mathbf{y}_{\op,i}) \approx \sigma^2_{\post,j}$ for all $i \in \mathcal{I}_j$, where $\sigma^2_{\post,j}$ is the population subset-level expected posterior variance defined in Section~2.1. Under this approximation the per-examinee posterior variance and the routed-subset average coincide, so the empirical plug-in $\hat\sigma^2_{\post,j}$ defined in Section~3.2 as a sample average over the routed subset estimates the same population scalar that enters the closed-form correction.
\end{assumption}

\begin{assumption}[Joint Normality on the Routed Subset]\label{ass:joint_normal}
For the MLC estimators, the true latent trait $\theta_i$ and an imputed Plausible Value $\theta_i^{(s)}$ are approximately bivariate normal on the routed subset $\mathcal{I}_j$, with marginal mean $\mu_{\sub,j}$, marginal variance $\sigma^2_{\sub,j}$, and correlation $\rho_j = 1 - \sigma^2_{\post,j} / \sigma^2_{\sub,j}$.
\end{assumption}

\paragraph{Remark on Approximation~\ref{ass:joint_normal} under selective routing.}
Approximation~\ref{ass:joint_normal} holds whenever routing leaves $(\theta_i, \theta_i^{(s)})$ bivariate normal on the routed subset, as under missing-completely-at-random (MCAR) routing. The Fisher-information-targeted routing of Section~3 selects on $\bar\theta_i$ and reweights the routed-subset distribution away from bivariate normality. More generally, practical administration designs cannot be expected to produce Gaussian routed subsets. The approximation is instead supported by its empirical plug-ins $(\hat\rho_j, \hat\mu_{\sub,j}, \hat\sigma^2_{\sub,j}, \hat\sigma^2_{\post,j})$, which are computed on the realized routed subset and absorb the location and range-restriction effects that routing induces. The simulations in Section~3 indicate that the residual dependence on distributional shape is benign in the studied conditions.

Under these approximations, the implied marginal response function on which the local regressions converge can be evaluated in closed form via Gaussian convolution, leading to the deterministic attenuations of Sections~2.3.2 and 2.3.3 below. All proofs are deferred to Appendix~\ref{app:proofs}, where they appear under headings that match the proposition names.

\subsubsection{Asymptotic Bias in PELC: Probit Convolution}

PELC uses the EAP $\bar\theta_i$ as a fixed predictor. Under Approximations~\ref{ass:posterior_normal}--\ref{ass:homosc}, the posterior residual $e_i = \theta_i - \bar\theta_i$ satisfies $e_i \mid \bar\theta_i \stackrel{\cdot}{\sim} \mathcal{N}(0, \sigma^2_{\post,j})$ on the routed subset. The marginal response function conditional on the EAP is obtained by integrating the true response function over this distribution:
\begin{equation}
    P(Y_{ij} = 1 \mid \bar{\theta}_i) \;\approx\; \int \Phi\left(\frac{a_j \bar{\theta}_i + d_j + a_j e_i}{D}\right) \phi(e_i; 0, \sigma^2_{\post,j}) \, de_i.
\end{equation}
Applying the standard Gaussian--probit convolution identity, $\int \Phi(A + Bx)\phi(x; 0, \sigma^2)dx = \Phi(A/\sqrt{1+B^2\sigma^2})$ \citep{lord1968}, we obtain:
\begin{equation}
    P(Y_{ij} = 1 \mid \bar{\theta}_i) \approx \Phi\left( \frac{(a_j \bar{\theta}_i + d_j) / D}{\sqrt{1 + (a_j/D)^2 \sigma^2_{\post,j}}} \right).
\end{equation}
Converting back to the logistic metric reveals the structural PELC attenuation. Because the EAP acts as the conditional mean, this attenuation depends only on the subset-average posterior variance and is independent of $\sigma^2_{\sub,j}$:
\begin{align}
    E[\hat{a}_{\PELC}] &\approx \frac{a_j}{\sqrt{1 + \frac{a_j^2}{D^2} \sigma^2_{\post,j}}}, \label{eq:pelc_slope} \\
    E[\hat{d}_{\PELC}] &\approx \frac{d_j}{\sqrt{1 + \frac{a_j^2}{D^2} \sigma^2_{\post,j}}}. \label{eq:pelc_intercept}
\end{align}

\begin{proposition}[PELC Attenuation]\label{prop:pelc_atten}
Under Approximations~\ref{ass:normal_ogive}--\ref{ass:homosc}, the population limits of the PELC estimators are given by the right-hand sides of Equations~\eqref{eq:pelc_slope} and~\eqref{eq:pelc_intercept}.
\end{proposition}

\subsubsection{Asymptotic Bias in MLC: Double Attenuation}

Unlike the EAP, individual Plausible Values ($\theta_i^{(s)}$) carry injected noise to recover the full marginal variance. Because the local regression sees only the routed subset, estimation is governed by the restricted trait distribution of that subset. Let $\bar{\theta}_{\sub,j}^{(s)} = \frac{1}{N_j}\sum_{i \in \mathcal{I}_j} \theta_i^{(s)}$ denote the imputation-$s$ subset mean. The empirical subset mean ($\hat\mu_{\sub,j}$) and variance ($\hat\sigma^2_{\sub,j}$) are obtained by averaging across imputations:
\begin{equation}
    \hat\mu_{\sub,j} = \frac{1}{S}\sum_{s=1}^S \bar{\theta}_{\sub,j}^{(s)}, \quad \hat\sigma^2_{\sub,j} = \frac{1}{S}\sum_{s=1}^S \left[ \frac{1}{N_j - 1} \sum_{i \in \mathcal{I}_j} \left( \theta_i^{(s)} - \bar{\theta}_{\sub,j}^{(s)} \right)^2 \right].
\end{equation}
These converge in probability to $\mu_{\sub,j}$ and $\sigma^2_{\sub,j}$ under the joint asymptotics of Section~2.5.

Under the homoscedastic approximation, the local reliability of the Plausible Values is the ratio of true to total subset variance, $\rho_j = (\sigma^2_{\sub,j} - \sigma^2_{\post,j}) / \sigma^2_{\sub,j} = 1 - \sigma^2_{\post,j} / \sigma^2_{\sub,j}$, with empirical analogue $\hat\rho_j = 1 - \hat\sigma^2_{\post,j}/\hat\sigma^2_{\sub,j}$. This reliability adapts to the restricted range of the sample: highly targeted subgroups exhibit lower local reliability than the global population.

Treating a PV as a fixed covariate induces non-classical errors-in-variables (EIV) attenuation. Because $\theta_i$ and $\theta_i^{(s)}$ are independent draws from the same posterior, their correlation is exactly $\rho_j$ rather than the classical $\sqrt{\rho_j}$. Under joint normality, the conditional expectation of the true trait given the PV shrinks toward $\mu_{\sub,j}$:
\begin{equation}
    E[\theta_i \mid \theta_i^{(s)}] = \rho_j \theta_i^{(s)} + (1-\rho_j)\mu_{\sub,j},
\end{equation}
with residual variance $\text{Var}(\theta_i \mid \theta_i^{(s)}) = \sigma^2_{\sub,j}(1 - \rho_j^2)$. Writing $\theta_i = \rho_j \theta_i^{(s)} + (1-\rho_j)\mu_{\sub,j} + u_i$ with $u_i \sim \mathcal{N}\left(0, \sigma^2_{\sub,j}(1-\rho_j^2)\right)$, the marginal response conditional on the MLC predictor is:
\begin{equation}
    P(Y_{ij} = 1 \mid \theta_i^{(s)}) \;\approx\; \int \Phi\!\left(\frac{a_j \left(\rho_j \theta_i^{(s)} + (1-\rho_j)\mu_{\sub,j}\right) + d_j + a_j u_i}{D}\right) \phi\left(u_i;\, 0,\, \sigma^2_{\sub,j}(1-\rho_j^2)\right)\, du_i.
\end{equation}
The convolution reveals a \textit{double attenuation} from two mechanisms: a multiplicative regression-to-the-mean factor $\rho_j$, reflecting that $\theta_i$ and $\theta_i^{(s)}$ are exchangeable posterior draws (the OLS regression of $\theta_i$ on $\theta_i^{(s)}$ has slope $\rho_j$ and passes through $(\mu_{\sub,j}, \mu_{\sub,j})$); and a divisive probit-convolution factor over residual EIV variance $\sigma^2_{\sub,j}(1-\rho_j^2)$, analogous to PELC but with EIV variance rather than $\sigma^2_{\post,j}$:
\begin{align}
    E[\hat{a}_{\MLC}] &\approx \frac{a_j \cdot \rho_j}{\sqrt{1 + \frac{a_j^2}{D^2} \sigma^2_{\sub,j} (1 - \rho_j^2)}}, \label{eq:mlc_slope} \\
    E[\hat{d}_{\MLC}] &\approx \frac{d_j + a_j(1-\rho_j)\mu_{\sub,j}}{\sqrt{1 + \frac{a_j^2}{D^2} \sigma^2_{\sub,j} (1 - \rho_j^2)}}. \label{eq:mlc_intercept}
\end{align}
When the targeted-subgroup latent mean is non-zero ($\mu_{\sub,j} \neq 0$), the regression-to-the-mean shift $a_j(1-\rho_j)\mu_{\sub,j}$ enters the numerator of the intercept attenuation in Equation~\eqref{eq:mlc_intercept}, so that the corrected intercept must subtract this shift back out (Section~2.4).

\begin{proposition}[MLC Attenuation]\label{prop:mlc_atten}
Under Approximations~\ref{ass:normal_ogive}--\ref{ass:joint_normal}, the population limits of all three pooled MLC estimators (RR, MN, EC) are given by the right-hand sides of Equations~\eqref{eq:mlc_slope} and~\eqref{eq:mlc_intercept}.
\end{proposition}

\subsection{Analytic Bias Correction and Variance Scaling}

Because these attenuations are deterministic functions of the population nuisance quantities $(\rho_j, \mu_{\sub,j}, \sigma^2_{\sub,j}, \sigma^2_{\post,j})$ under the four approximations of Section~2.3.1, inverting these functions yields closed-form corrections applied at the empirical plug-ins. We organize the corrections multiplicatively, defining a scalar \textit{expansion factor} $\lambda$ for each estimator that maps the uncorrected slope onto its corrected value. The intercept correction follows mechanically.

\subsubsection{PELC Correction}

Let
\begin{equation}
\hat\lambda_{\PELC,j} \;=\; \left(1 \;-\; \frac{\hat a_{\PELC}^2}{D^2}\,\hat\sigma^2_{\post,j}\right)^{-1/2}
\label{eq:lambda_pelc}
\end{equation}
denote the PELC expansion factor. Inverting the probit-convolution attenuation of Section~2.3.2 at the empirical plug-ins gives
\begin{align}
\hat a_{\PELC,\corr} &= \hat\lambda_{\PELC,j}\, \hat a_{\PELC}, \\
\hat d_{\PELC,\corr} &= \hat\lambda_{\PELC,j}\, \hat d_{\PELC}.
\end{align}
Because PELC's attenuation factor is shared between slope and intercept, the same $\hat\lambda_{\PELC,j}$ scales both.

\subsubsection{MLC Correction}

For the MLC estimators, define the scaled residual-variance constant
\[
\hat c_j \;=\; \frac{\hat\sigma^2_{\sub,j}\,(1 - \hat\rho_j^2)}{D^2},
\]
and let
\begin{equation}
\hat\lambda_j \;=\; \left(\hat\rho_j^2 \;-\; \hat c_j\, \hat a_{\MLC}^2\right)^{-1/2}
\label{eq:lambda_mlc}
\end{equation}
denote the MLC expansion factor, which inverts both the regression-to-the-mean and the probit-convolution attenuation mechanisms identified in Section~2.3.3. The corrected slope and intercept are
\begin{align}
\hat a_{\MLC,\corr} &= \hat\lambda_j\, \hat a_{\MLC}, \label{eq:acorr_mlc} \\
\hat d_{\MLC,\corr} &= \hat\rho_j\, \hat\lambda_j\, \hat d_{\MLC} \;-\; \hat\lambda_j\,(1 - \hat\rho_j)\,\hat\mu_{\sub,j}\, \hat a_{\MLC}.
\label{eq:dcorr_mlc}
\end{align}
The intercept correction both rescales the attenuated intercept by $\hat\rho_j\hat\lambda_j$ and subtracts back the regression-to-the-mean translation $a_j(1-\rho_j)\mu_{\sub,j}$ identified in \eqref{eq:mlc_intercept}. The multiplicative form in \eqref{eq:dcorr_mlc} is written to remain finite and continuous at $\hat a_{\MLC}=0$, avoiding the instability of the algebraically equivalent $\hat d_{\MLC}/\hat a_{\MLC}$ form when slope estimates approach zero under sparse routing.

\begin{condition}[Admissibility]\label{cond:adm}
At the attenuated population coefficients $a_{\dagger,\PELC} = a_j / \sqrt{1 + a_j^2\sigma^2_{\post,j}/D^2}$ and $a_{\dagger,\MLC} = a_j \rho_j / \sqrt{1 + a_j^2 c_j}$ (with $c_j = \sigma^2_{\sub,j}(1 - \rho_j^2)/D^2$), the radicands of the inverse mappings, $1 - a_{\dagger,\PELC}^2 \sigma^2_{\post,j}/D^2$ for PELC and $\rho_j^2 - c_j a_{\dagger,\MLC}^2$ for MLC, are strictly positive.
\end{condition}

Direct algebraic substitution (see the proof of Proposition~\ref{prop:inverse}) shows that these population radicands are automatically strictly positive for any finite $a_j$ when $\rho_j > 0$. Empirical radicands evaluated at $\hat a_{\PELC}$ and $\hat a_{\MLC}$ may nonetheless be nonpositive in finite samples; Section~2.4.3 describes the finite-sample regularization.

\begin{proposition}[Inverse Mapping]\label{prop:inverse}
Evaluate the corrections \eqref{eq:lambda_pelc}--\eqref{eq:dcorr_mlc} at the population nuisance values (that is, with each hatted plug-in replaced by its population analogue). Under Approximations~\ref{ass:normal_ogive}--\ref{ass:homosc} and Condition~\ref{cond:adm} for PELC, and Approximations~\ref{ass:normal_ogive}--\ref{ass:joint_normal} and Condition~\ref{cond:adm} for MLC, these population-value corrections recover the generating parameters $(a_j, d_j)$ exactly. In the asymptotic setting of Section~2.5 ($N_j \to \infty$ with $J_{\op}$ and the operational bank fixed), the empirical corrections---evaluated at the imputed plug-ins $(\hat\rho_j, \hat\mu_{\sub,j}, \hat\sigma^2_{\sub,j}, \hat\sigma^2_{\post,j})$---recover $(a_j, d_j)$ in probability, provided these plug-ins converge to their population analogues\footnote{Because the $S$ operational-parameter draws are shared across all examinees, the between-operational-parameter component of $\hat\sigma^2_{\post,j}$ (and hence of $\hat\rho_j$) does not vanish as $N_j \to \infty$ with $S$ fixed; the plug-ins converge to their population analogues as $N_j$ and $S$ grow jointly, with the residual $O_p(S^{-1/2})$ contribution small whenever the between-operational-parameter term of Equation~\eqref{eq:totalvar_decomp} is small.} and Condition~\ref{cond:adm} holds in the limit. The exact recovery is therefore an asymptotic property; in finite samples the empirical radicands may be inadmissible and are regularized as described in Section~2.4.3.
\end{proposition}

\subsubsection{Admissibility and Finite-Sample Regularization}

The expansion factors \eqref{eq:lambda_pelc} and \eqref{eq:lambda_mlc} are real and finite only when the radicands are strictly positive: admissibility requires $\hat a_{\PELC}^2\, \hat\sigma^2_{\post,j} < D^2$ for PELC and $\hat\rho_j^2 > \hat c_j\, \hat a_{\MLC}^2$ for MLC. In finite samples with severe sparsity or range restriction, sampling variance may push the uncorrected discrimination estimates beyond these boundaries, yielding nonpositive radicands or arbitrarily large expansion factors. Severe targeting may additionally drive $\hat\sigma^2_{\sub,j}$ below $\hat\sigma^2_{\post,j}$ and produce a nonpositive empirical reliability $\hat\rho_j$. We regularize these boundary conditions by bounding the empirical reliability below at $0.001$, replacing $\hat\rho_j$ with $\max(\hat\rho_j, 0.001)$ in the correction, and capping the expansion factor at a fixed maximum $\lambda_{\max} = 3.0$. The cap truncates the correction when the local approximations degrade under extreme sparsity, at the cost of introducing residual attenuation in capped replications and a piecewise-defined sampling distribution near the admissibility boundary. We examine cap activation rates empirically in Section~3.4.6.

\subsubsection{Variance Scaling}

Because the uncorrected point estimates are shrunk toward zero, their covariance matrices (e.g., $\mathbf{T}_j$ for MLC-RR) are correspondingly compressed. To map standard errors back to the corrected scale, we apply a first-order multivariate delta-method correction. Let $\mathbf{J}$ denote the $2 \times 2$ Jacobian of the corrected estimator with respect to the uncorrected estimator. Because the corrected slope depends only on the uncorrected slope, $\mathbf{J}$ is lower-triangular with $J_{12} = 0$.

\paragraph{Jacobian on the smooth branch.} When the expansion factor is below the cap and the radicand is positive, the correction is the differentiable analytic inversion rather than the truncated constant of Section~2.4.3; we call this the \emph{smooth branch}. Differentiating \eqref{eq:acorr_mlc} and \eqref{eq:dcorr_mlc} at the empirical plug-ins gives
\begin{align}
J_{11}^{\text{smooth}} &= \frac{\partial \hat a_{\MLC,\corr}}{\partial \hat a_{\MLC}} \;=\; \hat\rho_j^{\,2}\, \hat\lambda_j^{\,3}, \\
J_{22}^{\text{smooth}} &= \frac{\partial \hat d_{\MLC,\corr}}{\partial \hat d_{\MLC}} \;=\; \hat\rho_j\, \hat\lambda_j, \\
J_{21}^{\text{smooth}} &= \frac{\partial \hat d_{\MLC,\corr}}{\partial \hat a_{\MLC}} \;=\; \hat\lambda_j^{\,3} \hat\rho_j \!\left( \hat d_{\MLC}\, \hat a_{\MLC}\, \hat c_j \;-\; \hat\mu_{\sub,j}\, \hat\rho_j (1 - \hat\rho_j) \right).
\end{align}
The PELC analogues simplify to $J_{11}^{\text{smooth}} = \hat\lambda_{\PELC,j}^{\,3}$, $J_{22}^{\text{smooth}} = \hat\lambda_{\PELC,j}$, and $J_{21}^{\text{smooth}} = \hat d_{\PELC}\, \hat a_{\PELC}\, \hat\lambda_{\PELC,j}^{\,3}\,(\hat\sigma^2_{\post,j}/D^2)$.

\paragraph{Jacobian on the capped branch.} When the expansion factor is truncated at $\lambda_{\max}$, the smooth-branch chain rule no longer applies: the transformation becomes linear in $(\hat a_{\MLC}, \hat d_{\MLC})$, the chain-rule terms through $\partial \hat\lambda_j / \partial \hat a_{\MLC}$ vanish, and the Jacobian reduces to that of the resulting constant-multiplier map. The explicit piecewise-linear matrix forms for both PELC and MLC are given in Supplement~S6.4. The operational implementation evaluates $\mathbf{J}$ piecewise; using the smooth formulas in capped replications would substantially overstate the propagated variance.

The corrected $2 \times 2$ covariance matrix is then computed by sandwiching the uncorrected covariance matrix with the appropriate Jacobian:
\begin{equation}
    \widehat{\mathbf{V}}_{\corr} = \mathbf{J}\, \widehat{\mathbf{V}}_{\uncorr}\, \mathbf{J}^\intercal.
\end{equation}
The variance adjustment conditions on the empirical nuisance plug-ins $(\hat\rho_j, \hat\mu_{\sub,j}, \hat\sigma^2_{\sub,j}, \hat\sigma^2_{\post,j})$ as fixed, so first-order uncertainty in those quantities and their cross-covariance with the uncorrected estimate is not propagated. Section~4.3 discusses a sandwich-form refinement that addresses these contributions.

\begin{proposition}[Conditional Smooth-Branch Delta-Method Covariance]\label{prop:jacobian}
On the smooth branch (Approximations~\ref{ass:normal_ogive}--\ref{ass:joint_normal} and Condition~\ref{cond:adm} hold at the empirical plug-ins), the propagated covariance $\widehat{\mathbf{V}}_{\corr} = \mathbf{J}\,\widehat{\mathbf{V}}_{\uncorr}\,\mathbf{J}^{\intercal}$ is the first-order delta-method approximation conditional on the realized routed subset and the empirical attenuation quantities $(\hat\rho_j, \hat\mu_{\sub,j}, \hat\sigma^2_{\sub,j}, \hat\sigma^2_{\post,j})$, with $\widehat{\mathbf{V}}_{\uncorr}$ interpreted as the variance estimate of the pooled uncorrected coefficient marginally over the Plausible-Values draws (e.g., $\mathbf{T}_j$ for MLC-RR). Unconditional refinement is discussed in Section~4.3.
\end{proposition}

\subsection{Asymptotic Behavior of the Corrected Estimator}

The asymptotic setting we consider holds the operational test length $J_{\op}$ and item bank fixed (with the operational parameters estimated from the calibration sample and their uncertainty propagated through the Plausible Values) and lets the pilot sample $N_j$ grow under the fixed routing rule. In this regime the corrected estimator converges in probability to $\boldsymbol{\eta}_j$ under the stated approximations (Proposition~\ref{prop:inverse}). A defining feature of the construction is that the correction is required even in the limit: $N_{\op}\to\infty$ alone does not send $\sigma^2_{\post,j}$ to zero, because with $J_{\op}$ fixed only the between-operational-parameter term of \eqref{eq:totalvar_decomp} vanishes while the within-operational-parameter term governed by $J_{\op}$ has a strictly positive floor. The attenuation is therefore structural rather than a small-sample effect, and an uncorrected plug-in such as Method A retains its bias as $N_j$ grows.

Modularization itself carries an information cost relative to a concurrent full-information calibration that re-estimates the operational scale jointly with the pilot: blocking the pilot-to-trait feedback channel inflates the asymptotic variance of the pilot parameters. Supplement~S7 shows this loss is governed by the ratio of the pilot item's trait information to the operational module's, and is negligible in pretesting applications, where the operational module dominates trait estimation.

While the PELC point estimate is analytically correctable, treating the EAP as an error-free covariate underestimates the total variance, yielding anti-conservative standard errors at moderate-to-large $N_j$ once local-regression variability no longer dominates. Posterior integration via MLC propagates this uncertainty and is preferred for inferential applications. 

\section{Monte Carlo Simulation Study}

The simulation study evaluates the local item calibration estimators with four objectives: (1) verify that the closed-form correction recovers approximately unbiased $(a_j, d_j)$ under the stated approximations; (2) characterize the empirical 95\% interval coverage of the Jacobian-scaled covariance; (3) compare the corrected estimator against the modular online-calibration comparators across sparsity and routing conditions; and (4) diagnose the analytic standard errors against a population delta-method benchmark. Each replication calibrates a \emph{single} pilot item under the unidimensional 2PL. Mirroring the modular construction of Section~2, the operational item parameters are calibrated on one examinee sample, while a disjoint sample is administered the same operational items, with a routed subset additionally answering the pilot item.

\subsection{Experimental Design and Data Generation}

The simulation conditions cross six factors, which act on the pilot calibration through three channels. The operational test length $J_{\op} \in \{25, 50, 100\}$ controls the within-operational-parameter component of the posterior variance \eqref{eq:totalvar_decomp}, its dominant and non-vanishing part, while the operational calibration sample size $N_{\op} \in \{250, 1{,}000\}$ controls only its smaller between-operational-parameter component, through the width of the operational-parameter posterior. The pilot sample size $N_{\pilot} \in \{50, 100, 250, 500, 1{,}000\}$ and the routing mechanism (MCAR or Adaptive) jointly determine the routed subset: its size and, under Adaptive routing, its restricted trait range. The pilot item's own generating parameters, crossed as $a_{\true} \in \{0.8, 1.5\}$ with $d_{\true} \in \{0, -1\}$, fix the four pilot-parameter cells whose attenuation the correction must recover. The remaining factor, the number of Plausible Values $S \in \{5, 10, 20, 30, 50, 75, 100\}$, is a property of the estimator rather than the data. Main-text figures report the headline cell $(a_{\true}, d_{\true}) = (1.5, -1.0)$, representative of highly discriminating pilot items and the conditions in which the correction is most consequential; the other three cells appear in the supplementary materials.

Data are generated under the two-cohort design described above, which mirrors operational deployment: the bank is calibrated before pilot data arrive and then held fixed. For each combination of $(J_{\op}, N_{\op}, a_{\true}, d_{\true}, \text{routing})$ a single operational bank is drawn ($a_{\op} \sim \mathrm{Lognormal}(0, 0.25)$, with $0$ and $0.25$ the mean and standard deviation of $\log a_{\op}$; $d_{\op} \sim \mathcal{N}(0, 1)$) and reused as the data-generating bank across all $N_{\pilot}$ levels and all $R = 1{,}000$ replications.
Because a fresh bank is drawn for each such combination, comparisons across $J_{\op}$ and $N_{\op}$ cells reflect both the manipulated factor and the particular bank realization; comparisons across $N_{\pilot}$ levels and across estimators within a cell are unaffected. Within a replication, two independent examinee samples are drawn from $\mathcal{N}(0,1)$. The \emph{calibration cohort} of $N_{\op}$ examinees is administered the full $J_{\op}$-item operational test, from which the operational item parameters and their sampling covariance are estimated by marginal maximum likelihood. The \emph{pilot-administration cohort} of $5{,}000$ examinees is administered the same operational items, scored at the fixed calibration estimates with its operational responses never re-entering the bank, and its routed subset is additionally administered the single pilot item. All estimators within a replication are computed from these same responses, so the comparisons are paired.

Two routing mechanisms assign the pilot item within the pilot cohort, with the routed-subset size fixed at $N_{\pilot}$ under both. MCAR samples $N_{\pilot}$ examinees uniformly at random. Adaptive routing operates in two stages: each examinee in the pilot cohort is first admitted to a routing pool independently with probability equal to the max-normalized Fisher-information weight of the pilot item, $\pi_i = w_i / \max_k w_k$ with $w_i = a_{\true}^2\, P_i(1 - P_i)$ and $P_i$ evaluated at the examinee's operational EAP; the routed subset of exactly $N_{\pilot}$ examinees is then drawn uniformly without replacement from the admitted pool (supplemented by a uniform draw from the remainder of the cohort in the rare replications where fewer than $N_{\pilot}$ are admitted, which the simulation flags). Marginally, inclusion is approximately proportional to $w_i$, and the population routed-subset trait density is proportional to $\phi(\theta)\,\pi(\theta)$, the density against which the delta-method benchmark of Section~3.3 integrates. The routing EAP is computed at the generating operational item parameters, representing a deployed routing system with a well-calibrated bank; operational calibration uncertainty enters the estimators rather than the routing rule.

\subsection{Estimator Implementation}

For each replication we compute three estimator families: the modular comparators of Section~1.2, the corrected PELC and MLC estimators of Section~2, and unpenalized parallels that isolate Firth's contribution. All estimators are implemented in R \citep{R2026}.

\paragraph{Modular comparators.} Comparators are restricted to estimators that share the modular constraint motivating the framework (trait estimates constructed without the focal pilot's responses), so MEM and joint full-information calibration, which restore the pilot-to-trait feedback channel, are excluded by design; the information cost of this constraint relative to full information is quantified analytically in Supplement~S7. Three estimators sharing this modular structure serve as comparators. Method A \citep{stocking1988} regresses the pilot responses on the operational maximum-likelihood ability estimate $\hat\theta_i^{O}$ as a fixed predictor, with the standard error taken from the logistic-regression covariance. OEM \citep{wainermislevy1990} maximizes the modular marginal likelihood \eqref{eq:mod_likelihood} for the pilot item, integrating against each examinee's operational posterior held fixed at the operational MLE of the bank parameters, with the standard error from the inverse observed information (numerical Hessian) at the optimum. The per-examinee operational posteriors that serve as OEM's quadrature weights are evaluated on $61$ equally spaced nodes spanning $[-6, 6]$ (the grid used for all posterior computations in the simulation), and the marginal likelihood is maximized by quasi-Newton (BFGS) iteration.
FFMLE-M-Method A \citep{chenwang2016} applies the functional measurement-error correction of \citet{stefanskicarroll1985} to a Method A fit (Individual scheme, using the inverse operational Fisher information at each examinee's MLE; the Mean-scheme variant is evaluated in Supplement~S5.2). Because \citet{chenwang2016} prescribe no standard error, we report the second-stage logistic-regression covariance, treating the corrected ability as fixed.
Examinees with perfect or zero operational raw scores, for whom the operational MLE diverges and $\hat\theta_i^{O}$ and $\Sigma_i$ are undefined, are dropped from the Method~A and FFMLE regressions; OEM retains all routed examinees, because the operational posterior remains proper under the trait prior even for extreme response patterns. Finite operational MLEs are computed by damped Newton--Raphson on $[-6, 6]$ with a grid-search fallback.

\paragraph{Operational-module imputation.} The operational items are calibrated on the calibration cohort by marginal maximum likelihood via the EM algorithm (\texttt{mirt}; \citealp{chalmers2012}), with the parameter covariance taken from the observed information computed by the Oakes identity \citep{oakes1999}, and their sampling distribution is approximated as multivariate Gaussian at the MLE on the raw $(a_{\op}, d_{\op})$ scale.
Drawing $S = 100$ operational-parameter vectors $\boldsymbol{\eta}_{\op}^{(s)}$ from this approximation and computing, for each draw, every pilot-cohort examinee's operational posterior yields $S$ Plausible Values $\theta_i^{(s)}$ together with posterior means $m_i^{(s)} = E(\theta_i \mid \mathbf{y}_{\op,i}, \boldsymbol{\eta}_{\op}^{(s)})$ and variances $v_i^{(s)} = \mathrm{Var}(\theta_i \mid \mathbf{y}_{\op,i}, \boldsymbol{\eta}_{\op}^{(s)})$. The marginal posterior variance follows from the law-of-total-variance plug-in
\begin{equation}
\widehat{\mathrm{Var}}(\theta_i \mid \mathbf{y}_{\op,i}, \mathbf{Y}_{\op}) \;=\;
\frac{1}{S}\sum_{s=1}^{S} v_i^{(s)} \;+\;
\frac{1}{S-1}\sum_{s=1}^{S} (m_i^{(s)} - \bar m_i)^2,
\qquad \bar m_i = \frac{1}{S}\sum_{s=1}^{S} m_i^{(s)},
\label{eq:postvar_estimator}
\end{equation}
whose two terms estimate the within- and between-operational-parameter components of \eqref{eq:totalvar_decomp}, with $\hat\sigma^2_{\post,j} = N_j^{-1}\sum_{i \in \mathcal{I}_j} \widehat{\mathrm{Var}}(\theta_i \mid \mathbf{y}_{\op,i}, \mathbf{Y}_{\op})$. Smaller $S$ are evaluated as nested subsets of these $100$ draws.

\paragraph{Local pilot calibration.} The routed pilot responses are regressed on the EAP $\bar\theta_i$ for PELC (operationalized as the parameter-averaged posterior mean $\bar m_i$ of Equation~\eqref{eq:postvar_estimator}) and on each of the $S$ Plausible Values for the three MLC variants (RR, MN, EC), in both unpenalized and Firth-penalized form (\texttt{stats::glm} and \texttt{brglm2}; \citealp{kosmidis2020}). Section~3.4.4 isolates the Firth-specific contribution.

\paragraph{Closed-form correction and truncation.} The closed-form inversions and Jacobian scaling of Section~2.4 are applied to the pooled quantities (pool-then-correct): for RR and MN the pooled uncorrected point estimate and pooled covariance matrix, and for EC the empirical mean and covariance of the stochastic draws, are mapped to the corrected scale, with the subset moments computed from the same Plausible Values, the empirical reliability $\hat\rho_j$ floored at $0.001$, and the expansion factor capped at $\lambda_{\max} = 3.0$ (using the capped-branch Jacobian when the cap is active). Because the correction is nonlinear, correcting each imputation and then pooling would differ in finite samples; pooling first applies the inversion at the less variable input. All estimators then truncate the reported point estimates at $|\hat a| \le 5$ and $|\hat d| \le 5$. Non-converged \texttt{glm} fits under separation are likewise retained and truncated.

\subsection{Evaluation Criteria and Population Delta-Method Benchmark}

Performance is summarized by the mean and median bias of the point estimate, the mean squared error (MSE), the empirical sampling SD across replications, the analytic SE (its median across replications), and the empirical coverage of nominal 95\% confidence intervals. Mean and median bias are both reported because the nonlinear correction can right-skew the sampling distribution at small $N_{\pilot}$: a translated distribution registers as mean bias, whereas one centered near the truth with a heavy upper tail registers as mean bias exceeding median bias. Median-unbiasedness is the natural target for a nonlinearly transformed estimator \citep{kennepagui2017}.

To assess whether the PELC and MLC analytic SEs are correctly calibrated, we construct a population delta-method benchmark (\textit{PopDeltaSE}) for each cell. The attenuation corrections and their Jacobian (Section~2.4) are deterministic functions of the population nuisance quantities $(\mu_{\sub}, \sigma^2_{\sub}, \sigma^2_{\post}, \rho)$, which Propositions~\ref{prop:pelc_atten}--\ref{prop:mlc_atten} express as integrals over the routed-subset trait distribution. PopDeltaSE evaluates those integrals at the cell's data-generating values by Gauss--Hermite quadrature against the standard-normal trait density, reweighted by the routing probability $\pi(\theta)$ (which is constant across examinees under MCAR).

\subsection{Results}

\subsubsection{Bias, MSE, and Coverage at the Headline Cell}

\begin{figure}[t]
\centering
\includegraphics[width=\textwidth]{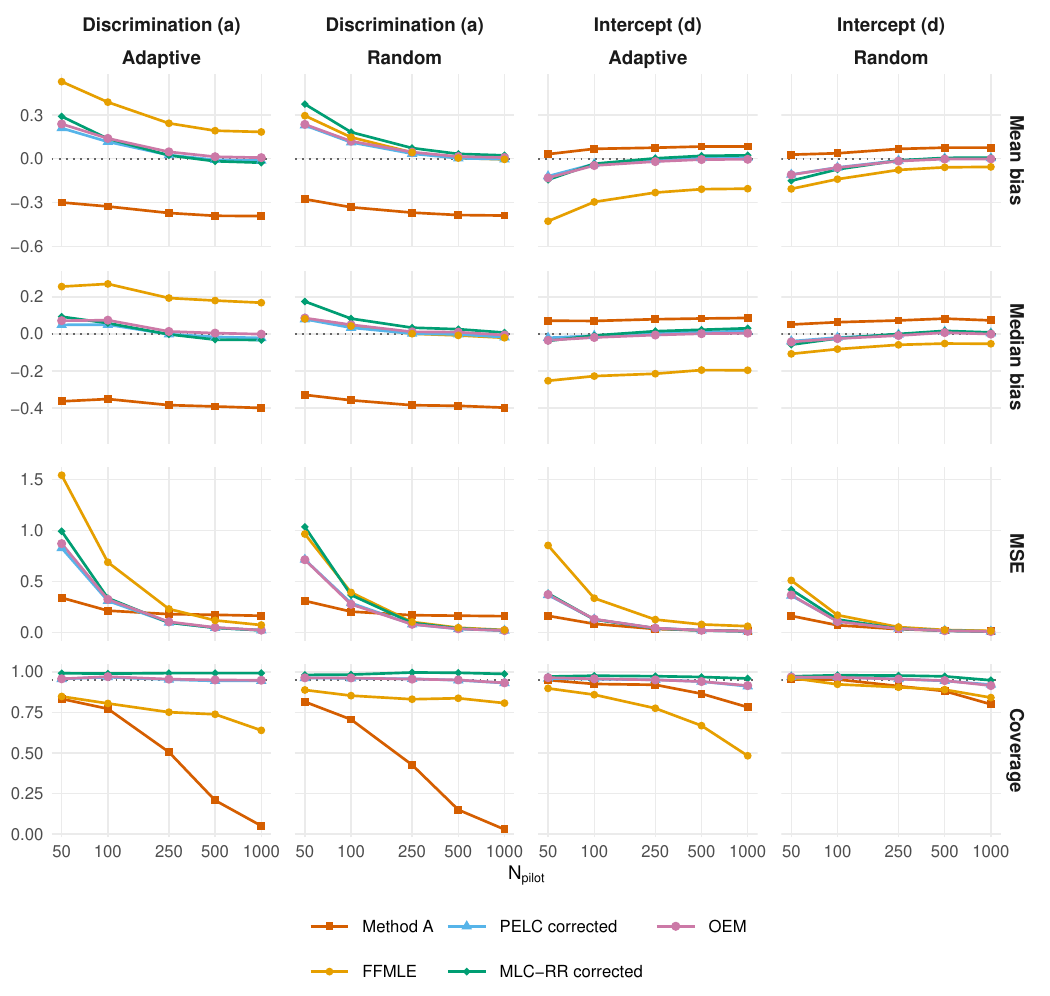}
\caption{Headline cell, both routing conditions: mean and median bias, mean squared error (MSE), and coverage of nominal 95\% confidence intervals for five modular estimators (Method A, FFMLE Individual, OEM, PELC corrected, and MLC-RR corrected), as a function of pilot sample size $N_{\pilot}$. Rows index the four evaluation metrics; columns index the cross of routing condition and item parameter. On-figure labels: Random denotes MCAR, and Adaptive denotes MAR via the Fisher-information targeting rule. PELC corrected and MLC-RR corrected are shown without Firth's penalty; their Firth-penalized variants appear in Figure~\ref{fig:gridpenalized}. Held fixed: $a_{\true} = 1.5$, $d_{\true} = -1.0$, $J_{\op} = 25$, $N_{\op} = 1{,}000$, $S = 100$.}
\label{fig:headline}
\end{figure}

Figure~\ref{fig:headline} shows five estimators across four metrics on each routing-by-parameter combination at the headline cell ($a_{\true} = 1.5$, $d_{\true} = -1.0$, $J_{\op} = 25$, $N_{\op} = 1{,}000$, $S = 100$). The estimators are Method A and FFMLE Individual (uncorrected modular comparators), OEM (modular comparator targeting the same population value as the corrected estimators by per-pilot quadrature), and PELC corrected and MLC-RR corrected (the proposed corrected estimators, shown without Firth's penalty for direct comparability with Method A and OEM; Firth-penalized variants in Section~3.4.4).

\paragraph{Discrimination bias.} Method A is structurally attenuated by approximately $-0.35$ to $-0.40$ in median bias under both routings and across all $N_{\pilot}$, with the attenuation persistent at $-0.40$ at $N_{\pilot} = 1{,}000$. The bias does not decay with $N_{\pilot}$, the textbook consequence of conditioning on a noisy proxy as if error-free. FFMLE Individual moves the bias in the opposite direction: under Adaptive routing, mean bias starts at approximately $+0.53$ at $N_{\pilot} = 50$ and decays only to $+0.19$ by $N_{\pilot} = 1{,}000$, with a substantial mean-versus-median gap ($+0.53$ versus $+0.25$ at $N_{\pilot} = 50$) indicating right-skewness. Under MCAR, FFMLE's bias is smaller and decays cleanly to near zero by $N_{\pilot} = 1{,}000$. This is the setting for which \citet{chenwang2016} developed FFMLE-M-Method A: their measurement-error framework models the operational MLE error $\xi_i = \hat\theta_i^O - \theta_i$ as multivariate normal, conditionally zero-mean given $\theta_i$, and independent of the pilot response $y_{ij}$, with the multivariate-CAT extension generalizing the error covariance to a per-examinee $\Sigma_i = I(\theta_i)^{-1}$ varying with $\theta_i$ while preserving the conditional centering (their Section~3.3.1, Eq.~11, inheriting the FFMLE independence assumption $\xi_i \perp y_{ij}$, their $\varepsilon_i$, from their Section~3.2). Their simulation administered pilot items by random selection (their Section~4.3), and adaptive routing of pilot items is left for future investigation (their Section~6). The Adaptive-routing comparison reported here extends the evaluation set rather than contesting FFMLE's established performance under random routing; it is informative because online-calibration deployments increasingly route pilot items adaptively, and the two corrections handle the routing-induced distributional shifts via structurally different mechanisms.

When the focal pilot item is routed adaptively (examinee $i$ admitted to $\mathcal{I}_j$ with probability $\pi(\bar\theta_i)$ depending on the operational EAP), the centering property fails on $\mathcal{I}_j$. Because the operational EAP and the operational MLE are both deterministic functions of the same operational responses, selection on $\bar\theta_i$ reweights the joint distribution of $(\theta_i, \xi_i)$ among admitted examinees, where $\xi_i$ is the MLE error of the FFMLE construction. Conditional on $\theta_i$ and selection $\{i \in \mathcal{I}_j\}$, the distribution of $\xi_i$ is reweighted by $\pi(\bar\theta_i)$, yielding $E[\xi_i \mid \theta_i, i \in \mathcal{I}_j] \neq 0$ in general. Because the FFMLE correction assumes zero-mean classical measurement error, applying it where this assumption is violated produces an overshoot in the opposite direction to Method A's attenuation. The proposed closed-form correction (Section~2.4) is less susceptible to this failure mode by structural design: its empirical plug-ins $(\hat\rho_j, \hat\sigma^2_{\sub,j}, \hat\sigma^2_{\post,j}, \hat\mu_{\sub,j})$ are computed on the realized routed subset and absorb the marginal trait location and range-restriction effects induced by routing, including the routing-shift term $-\hat\lambda_j(1-\hat\rho_j)\hat\mu_{\sub,j}\,\hat a_{\MLC}$ in the corrected intercept. The correction does not separately model residual conditional measurement-error distortions induced by selective routing; Supplement~S2.2 illustrates the form of this residual distortion, whereas Figures~\ref{fig:headline}--\ref{fig:gridfamily1} and the higher-order analysis in Supplement~S6 indicate that its effect on the corrected estimators is small in the simulated conditions. The empirical contrast in Figure~\ref{fig:headline} accords with this structural difference rather than being explained solely by finite-sample noise.

The PELC corrected, MLC-RR corrected, and OEM estimators behave nearly identically across both parameters and both routings. Under Adaptive routing at $N_{\pilot} = 50$, MLC-RR corrected reports a median bias of approximately $+0.10$ on $a$ and $-0.04$ on $d$; OEM reports $+0.07$ and $-0.04$; PELC corrected reports $+0.05$ and $-0.03$. By $N_{\pilot} = 1{,}000$ all three are within $\pm 0.04$ of zero on both parameters under both routings. The mean bias on $a$ is positive at small $N_{\pilot}$ for the corrected estimators (approximately $+0.29$ for MLC-RR corrected at $N_{\pilot} = 50$ Adaptive) and shrinks toward zero with $N_{\pilot}$, producing the MeanBias $>$ MedBias pattern that reflects right-skewness of the corrected sampling distribution. A second-order Taylor expansion of the correction $g(a) = a/\sqrt{\rho^2 - a^2 \sigma_e^2/D^2}$ around the attenuated population limit yields a positive second derivative on the smooth branch, which lifts the mean above $a_j$ proportionally to $\mathrm{Var}(\hat a_{j,\MLC}) = O(1/N_{\pilot})$. The intercept correction is linear in $\hat d_{\MLC}$ on the smooth branch, so the own-coordinate convexity term vanishes; the curvature entering through $\hat a_{\MLC}$ is evidently negligible in these conditions, and the intercept shows no analogous translation.

\paragraph{Coverage.} The most important result in Figure~\ref{fig:headline} is the bottom row, comparing coverage probability against the nominal 95\% target. Method A's coverage on the discrimination collapses with $N_{\pilot}$: from approximately $0.80$ at $N_{\pilot} = 50$ to $0.05$ at $N_{\pilot} = 1{,}000$ under Adaptive, and from $0.82$ to $0.03$ under MCAR. This combines asymptotic bias with shrinking SE: the $-0.40$ attenuation persists while the SE decays at the $\sqrt{N_{\pilot}}$ rate, placing the truth outside the interval in nearly every replication. FFMLE's coverage on $a$ degrades from $0.85$ at $N_{\pilot} = 50$ to $0.61$ at $N_{\pilot} = 1{,}000$ under Adaptive and stays near $0.83$--$0.89$ under MCAR; FFMLE's coverage on $d$ degrades from $0.89$ to $0.47$ at $N_{\pilot} = 1{,}000$ Adaptive, reflecting the persistent intercept residual. FFMLE's coverage should be read in light of the standard-error convention noted in Section~3.2: because \citet{chenwang2016} prescribe no standard error, the intervals evaluated here use the second-stage logistic-regression covariance, and alternative constructions could shift FFMLE's coverage in either direction. PELC corrected, MLC-RR corrected, and OEM hold coverage at or just above nominal throughout, with MLC-RR corrected ranging $0.96$--$0.99$ across the four routing-by-parameter combinations. The headline-cell parity between PELC corrected and MLC-RR corrected does not generalize: MLC-RR corrected's discrimination coverage stays at or above nominal everywhere evaluated, and its intercept coverage dips modestly below nominal only in extreme cells (with the worst case $0.93$ at $J_{\op} = 100$, $N_{\op} = 1{,}000$, $N_{\pilot} = 1{,}000$, $a_{\true} = 1.5$, $d_{\true} = 0$, Adaptive). PELC corrected's intercept coverage degrades more substantially in cells with small $N_{\op}$ and large $N_{\pilot}$ (with the worst case $0.76$ at $J_{\op} = 100$, $N_{\op} = 250$, $N_{\pilot} = 1{,}000$, $a_{\true} = 1.5$, $d_{\true} = 0$, Adaptive; within the headline-cell family $(1.5, -1.0)$, $0.81$ at the same $(J_{\op}, N_{\op}, N_{\pilot})$). This is consistent with the structural argument of Section~2.5: PELC treats the EAP as an error-free covariate, so its analytic SE captures the local-regression sampling variance and Jacobian propagation but omits the variability induced by the posterior uncertainty of the trait proxy, which MLC-RR propagates through Rubin's between-imputation component $\mathbf{B}_j$. When local-regression variance is small ($N_{\pilot}$ large), the omitted component accounts for a larger share of the total sampling variability. Section~4.2 returns to this when recommending MLC-RR over PELC for inferential applications.

The contrast between Method A and the corrected estimators on coverage is the central finding. The corrected methods trade slightly higher MSE at small $N_{\pilot}$ (Section~3.4.2) for nominal-to-conservative coverage that holds across the entire range studied. Method A's MSE advantage at small samples comes at the cost of a coverage failure that becomes catastrophic as the pilot sample grows.

\subsubsection{Bias-Variance Tradeoff in MSE}

Method A reports the lowest MSE on $a$ at small $N_{\pilot}$ despite substantial attenuation bias. At the headline cell ($J_{\op} = 25$, $N_{\op} = 1{,}000$), MSE on $a$ at $N_{\pilot} = 50$ Adaptive is $0.34$ for Method A, $0.83$ for PELC corrected, $0.99$ for MLC-RR corrected, $0.87$ for OEM, and $1.54$ for FFMLE Individual. This pattern reflects a classical bias-variance decomposition: Method A's squared bias ($\approx 0.13$) is smaller than the variance excess of the corrected methods, whose multiplicative correction Jacobian amplifies the input variance. Method A inherits only the local-regression sampling variance.

The MSE ordering reverses with $N_{\pilot}$. At $N_{\pilot} = 250$ Adaptive, MSE on $a$ is $0.18$ for Method A, $0.10$ for PELC corrected, $0.10$ for MLC-RR corrected, $0.11$ for OEM, and $0.23$ for FFMLE. By $N_{\pilot} = 1{,}000$, the corrected methods are at $0.022$--$0.024$ while Method A remains at $0.165$. Method A's MSE plateaus near its squared bias, whereas the corrected methods' MSE decays at the standard $1/N_{\pilot}$ rate. The crossover between Method A and the corrected estimators occurs near $N_{\pilot} = 100$--$250$ at the headline cell and shifts to larger $N_{\pilot}$ as $J_{\op}$ grows: the attenuation bias of Method A diminishes with longer operational tests, so its MSE plateau falls, and the corrected estimators need a larger pilot sample to drive their variance below it. At the cell most favorable to Method A ($J_{\op} = 100$, $N_{\op} = 250$), no crossover occurs within the simulated range $N_{\pilot} \le 1{,}000$ (Supplement~S8).

Method A's MSE advantage at small $N_{\pilot}$ is therefore real but comes at the cost of the coverage failures of Section~3.4.1. Pilot calibrations feed downstream operational scoring and adaptive routing; biased calibrations propagate as systematic errors there, while higher-variance unbiased calibrations average out across the item bank. We interpret the crossover as the cost of honest uncertainty propagation, not a recommendation to prefer Method A at small $N_{\pilot}$.

\subsubsection{Generalization Across the Design Grid}

\begin{figure}[t]
\centering
\includegraphics[width=\textwidth]{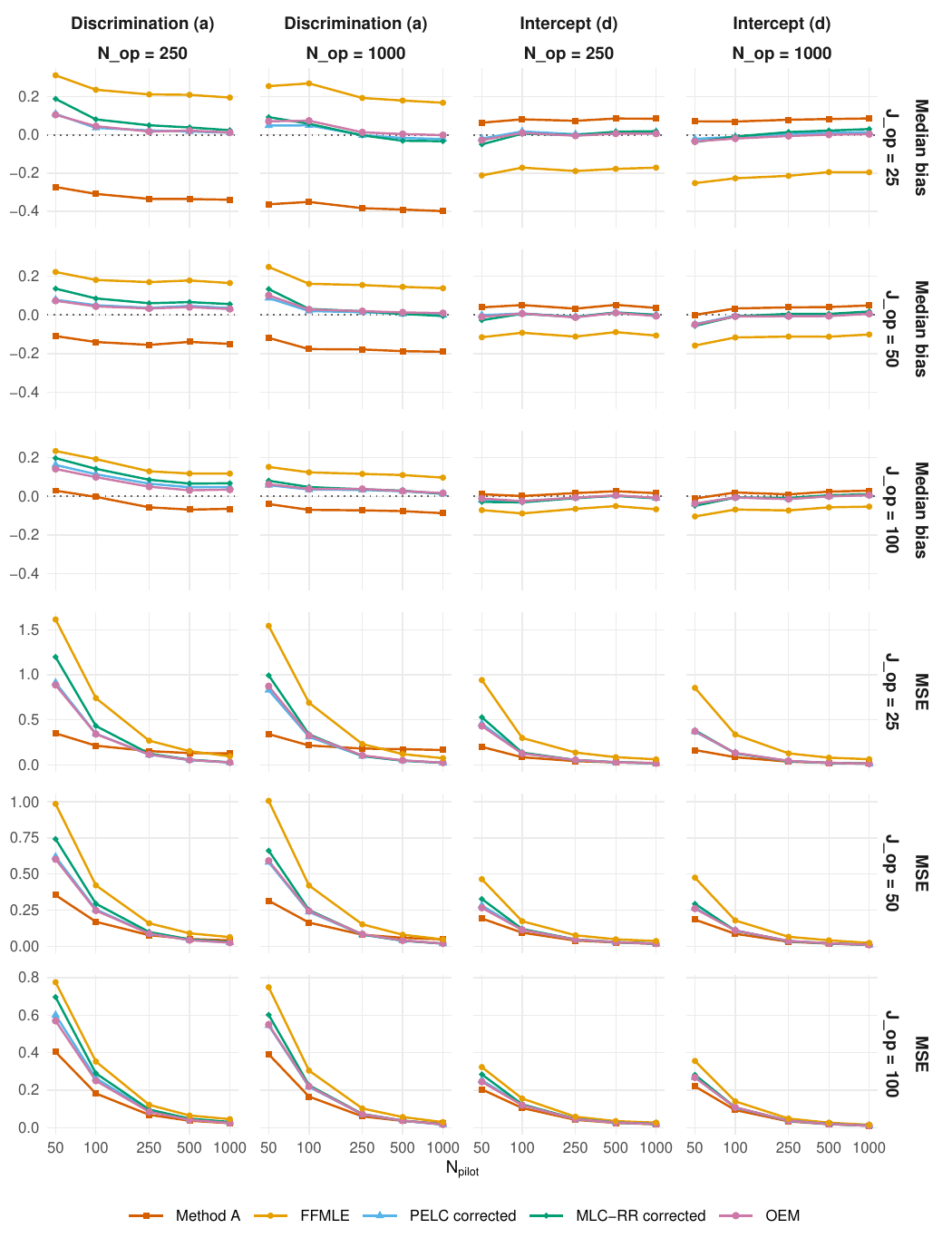}
\caption{Design-grid generalization, Adaptive routing: median bias and MSE for the five estimators of Figure~\ref{fig:headline}, as a function of pilot sample size $N_{\pilot}$. Rows are grouped by evaluation metric (median bias, then MSE), with $J_{\op} \in \{25, 50, 100\}$ varying within each metric block; columns are grouped by item parameter ($a$, then $d$), with $N_{\op} \in \{250, 1{,}000\}$ varying within each parameter block. Held fixed: $a_{\true} = 1.5$, $d_{\true} = -1.0$, $S = 100$. The MCAR analogue is Figure~S1 (Supplement~S2.1).}
\label{fig:gridfamily1}
\end{figure}

Figure~\ref{fig:gridfamily1} extends the comparison of Section~3.4.1 across the design grid: median bias and MSE for the same five estimators, broken out by $J_{\op} \in \{25, 50, 100\}$ (rows within each metric block) and $N_{\op} \in \{250, 1{,}000\}$ (columns within each parameter block) under Adaptive routing.

The qualitative patterns identified at the headline cell hold across the grid. Method A's median bias on $a$ shrinks with $J_{\op}$ (approximately $-0.40$ at $J_{\op} = 25$, $-0.17$ at $J_{\op} = 50$, and $-0.09$ at $J_{\op} = 100$) as operational measurement error diminishes with longer tests, but remains substantial at $J_{\op} = 100$, so structural attenuation persists at operationally realistic test lengths. FFMLE Individual's median bias on $a$ instead persists across $J_{\op}$: at $N_{\op} = 1{,}000$, the residual at $N_{\pilot} = 1{,}000$ is approximately $+0.18$, $+0.14$, and $+0.10$ at $J_{\op} = 25, 50, 100$, decaying only slowly. The PELC corrected, MLC-RR corrected, and OEM estimators stay within $\pm 0.20$ of zero on $a$ in every cell, with the largest residuals at $J_{\op} = 25$, $N_{\pilot} = 50$, where the correction's nonlinearity has the largest effect (Section~3.4.1). The corrected intercept's routing-shift term $-\hat\lambda_j(1-\hat\rho_j)\hat\mu_{\sub,j}\,\hat a_{\MLC}$ handles the $\hat\mu_{\sub,j} \neq 0$ that Adaptive routing induces.

The two $N_{\op}$ columns differ by a factor of four, yet the corrected methods' bias and MSE are nearly indistinguishable across them: at $J_{\op} = 25$, $N_{\pilot} = 50$ Adaptive, MLC-RR corrected has MSE on $a$ of $1.20$ at $N_{\op} = 250$ versus $0.99$ at $N_{\op} = 1{,}000$, with median bias $+0.19$ versus $+0.09$. This near-invariance follows from the decomposition in \eqref{eq:totalvar_decomp}: larger $N_{\op}$ shrinks only the between-operational-parameter component, while the dominant within-operational-parameter component governed by $J_{\op}$ continues to drive pilot-side estimation uncertainty. The Plausible-Values law-of-total-variance therefore leaves the corrected estimates stable across operational cohort sizes.

The MSE rows confirm the bias-variance tradeoff from the headline cell: Method A wins MSE on $a$ at small $N_{\pilot}$ in all six $(J_{\op}, N_{\op})$ combinations, with the gap to the corrected methods narrowing as $J_{\op}$ grows. At $J_{\op} = 100$, Method A's MSE on $a$ converges with MLC-RR corrected's by $N_{\pilot} = 500$ at $N_{\op} = 1{,}000$ (each $\approx 0.04$) and is overtaken by $N_{\pilot} = 1{,}000$; at $J_{\op} = 25$ the crossover comes earlier (between $N_{\pilot} = 100$ and $250$, depending on $N_{\op}$), as Method A's larger attenuation raises its MSE plateau in the short-test setting.

\subsubsection{Firth's Contribution to the Corrected Estimators}

\begin{figure}[t]
\centering
\includegraphics[width=\textwidth]{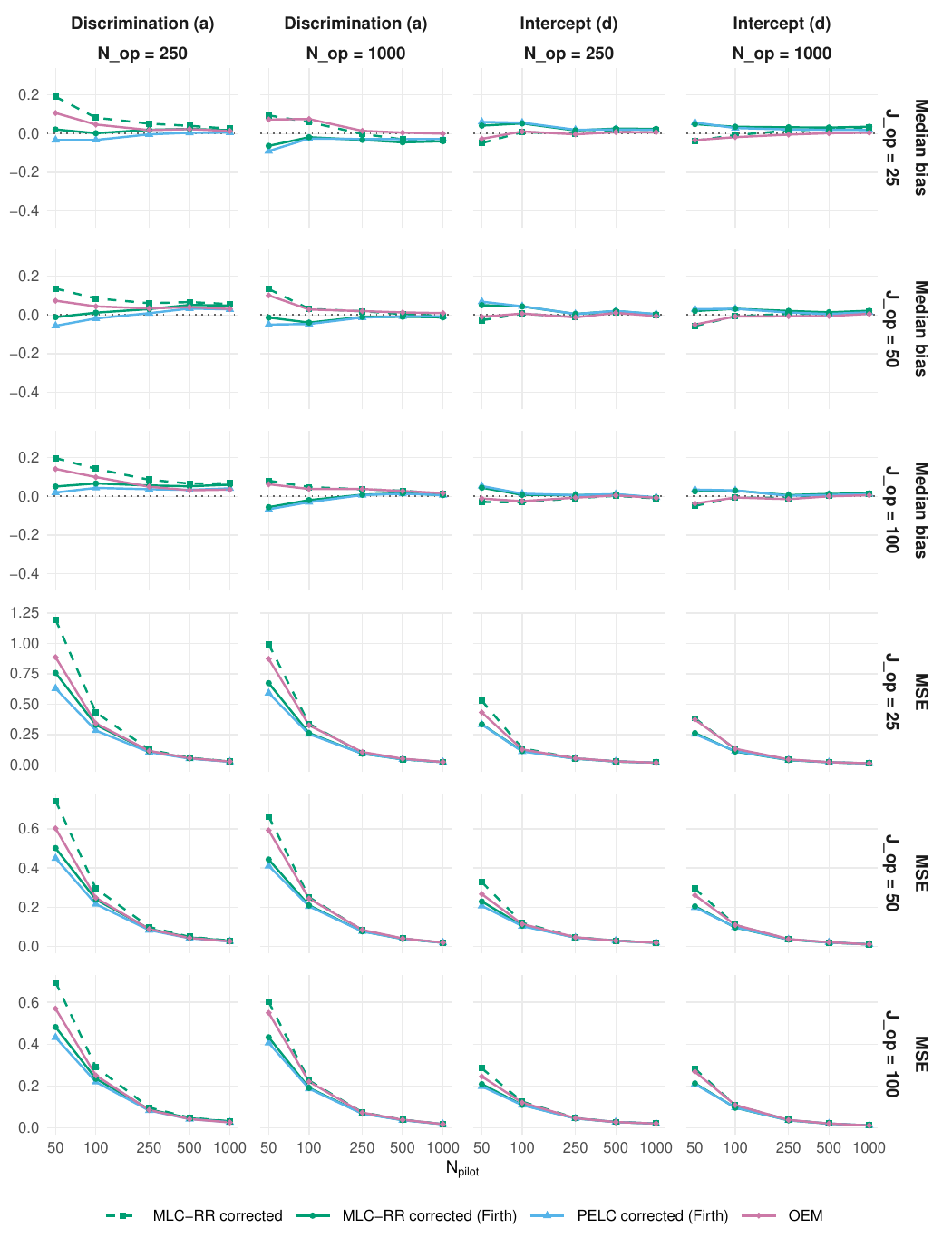}
\caption{Firth's contribution to the corrected estimators, Adaptive routing: median bias and MSE as a function of pilot sample size $N_{\pilot}$, for the MLC-RR corrected estimator without Firth's penalty (dashed) and with it (solid green), the Firth-penalized PELC corrected estimator (solid blue), and OEM (solid magenta). Rows are grouped by evaluation metric (median bias, then MSE), with $J_{\op} \in \{25, 50, 100\}$ varying within each metric block; columns are grouped by item parameter ($a$, then $d$), with $N_{\op} \in \{250, 1{,}000\}$ varying within each parameter block. Held fixed: $a_{\true} = 1.5$, $d_{\true} = -1.0$, $S = 100$. The MCAR analogue is Figure~S2 (Supplement~S2.1).}
\label{fig:gridpenalized}
\end{figure}

Figure~\ref{fig:gridpenalized} compares the unpenalized and Firth-penalized variants of the corrected MLC-RR estimator against PELC corrected (Firth) and OEM across the same $J_{\op} \times N_{\op}$ grid under Adaptive routing.

The unpenalized MLC-RR corrected estimator has visibly higher median bias on $a$ at small $N_{\pilot}$: at $J_{\op} = 25$, $N_{\op} = 250$, $N_{\pilot} = 50$ Adaptive, the unpenalized variant has median bias $+0.19$ on $a$ versus $+0.02$ for the Firth-penalized variant. The pattern matches the standard finite-sample bias of unpenalized MLE in sparse logistic regressions \citep{firth1993, kosmidis2020}. The MSE rows show a corresponding gap: for the same cell, the MSE on $a$ is $1.20$ for the unpenalized variant and $0.76$ for the Firth-penalized variant, a 37\% reduction. The gap closes by $N_{\pilot} = 250$, where the two variants are within 5\% of each other on MSE; at $N_{\pilot} = 1{,}000$ they are indistinguishable.

The PELC corrected (Firth) and OEM estimators track MLC-RR corrected (Firth) closely throughout. PELC corrected (Firth) has marginally lower MSE on $a$ at small $N_{\pilot}$ (e.g., $0.63$ versus $0.76$ at $J_{\op} = 25$, $N_{\op} = 250$, $N_{\pilot} = 50$). We attribute the lower MSE to PELC's single EAP-based proxy producing a less variable correction input, at the cost of not propagating posterior uncertainty into the standard error. OEM has slightly higher MSE on $a$ than the corrected estimators at small $N_{\pilot}$ (e.g., $0.89$ at the same cell), narrowing to indistinguishable performance by $N_{\pilot} = 250$. The closed-form correction therefore matches the per-pilot quadrature target it shares (Section~2; \citealp{wainermislevy1990}) without per-pilot numerical integration. Supplement~S2.1 replicates this pattern under MCAR; across both routings, Firth's contribution concentrates at small $N_{\pilot}$ and short $J_{\op}$, where sparse routed samples are most separation-prone.

\subsubsection{Standard Errors Are Approximately Calibrated to a Population Delta-Method Benchmark}

\begin{figure}[t]
\centering
\includegraphics[width=\textwidth]{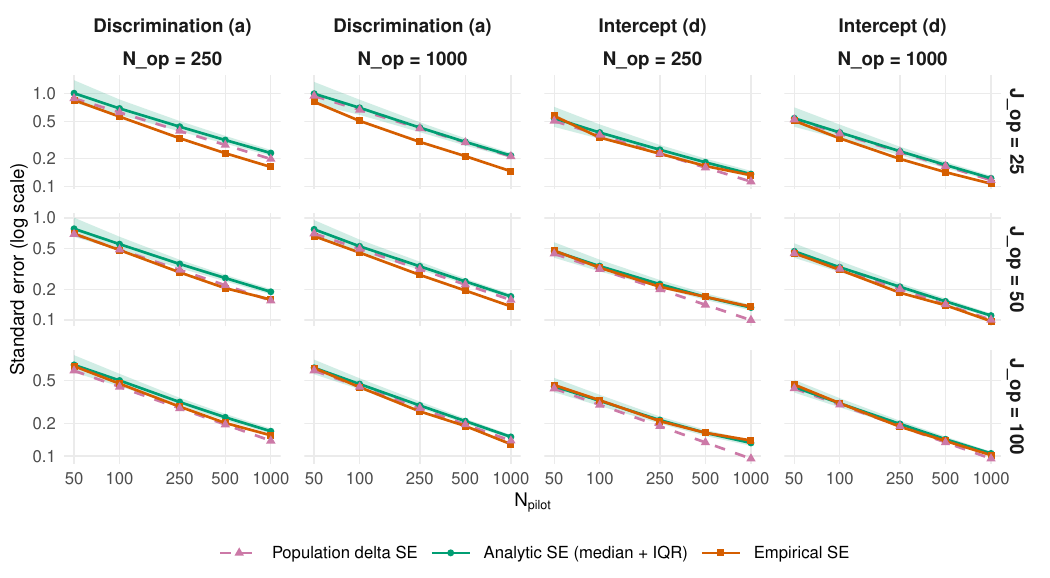}
\caption{Three-way standard-error comparison, Adaptive routing: the population delta-method benchmark (PopDeltaSE), the analytic SE (AnalyticSE; median across replications, with a shaded interquartile band), and the empirical sampling SD (EmpiricalSE) for the MLC-RR corrected (Firth) estimator, as a function of pilot sample size $N_{\pilot}$. Rows index $J_{\op}$; columns are grouped by $N_{\op}$, with item parameter ($a$, $d$) varying within each $N_{\op}$ block. All standard errors are plotted on a logarithmic scale. Held fixed: $a_{\true} = 1.5$, $d_{\true} = -1.0$, $S = 100$. The MCAR analogue is Figure~S3 (Supplement~S2.1).}
\label{fig:sethreeway}
\end{figure}

Figure~\ref{fig:sethreeway} compares the empirical sampling SD (EmpiricalSE), the analytic SE summarized by its median and interquartile-range (IQR) ribbon across replications (AnalyticSE), and the population delta-method benchmark (PopDeltaSE; Section~3.3) for the MLC-RR corrected (Firth) estimator across $J_{\op} \times N_{\op}$ under Adaptive routing.

The three quantities are broadly aligned across the grid, decreasing at similar rates with $N_{\pilot}$ and remaining within the same order of magnitude on the log scale, with two systematic discrepancies. AnalyticSE consistently exceeds EmpiricalSE by a roughly constant amount on the log scale, which yields the at-or-above-nominal coverage already noted. The smooth-branch Jacobian propagation evaluates the nuisance plug-ins as fixed, omitting their cross-covariance with the uncorrected estimate, which arises because both depend on the same operational Plausible Values. The net direction of the cross-terms is not determined a priori; Section~4.3 discusses this omission together with the supplementary empirical evidence that the omitted contributions are net negative across the design grid, with the cross-covariance term dominating the nuisance-variance term (Supplement~S6). PopDeltaSE tracks EmpiricalSE closely at small $N_{\pilot}$, but EmpiricalSE falls progressively below PopDeltaSE as $N_{\pilot}$ grows, with the gap larger at small $J_{\op}$.

The second pattern does not violate any lower bound, because PopDeltaSE is not a Cram\'er--Rao bound for the corrected estimator. It is the smooth-branch delta-method propagation of the local-regression sampling variance through the population-level Jacobian at population nuisance values, and omits three properties of the actual estimator: the finite-sample bias of Section~3.4.1 (Cram\'er--Rao does not apply to biased estimators), the piecewise-linear capped-branch transformation when $\hat\lambda_j = \lambda_{\max}$, and the joint variability of the empirical attenuation quantities---equivalently, the three nuisance plug-ins ($\hat\mu_{\sub,j}$, $\hat\sigma^2_{\sub,j}$, $\hat\sigma^2_{\post,j}$) together with the derived reliability $\hat\rho_j = 1 - \hat\sigma^2_{\post,j}/\hat\sigma^2_{\sub,j}$, which all share dependence on the operational Plausible Values. Section~4.3 discusses the sandwich-form refinement addressing the third source.

\subsubsection{Supplementary Findings}

Five additional analyses are documented in the supplementary materials. The three MLC pooling rules (RR, MN, EC) yield nearly identical point estimates and coverage (Supplement~S3.2): RR and MN coincide on the point estimate by construction, differing only in the variance penalty, and in the full simulation results stochastic EC differs from RR by less than $0.03$ on both $|\mathrm{MedBias}_a|$ and $\mathrm{MSE}_a$ in every cell evaluated. Sensitivity to $S$ is small for $S \geq 20$, with the larger gains from $S = 5$ to $S = 20$ reflecting the $S^{-1}$ component of Rubin's total-variance formula (Supplement~S3.1). Cap activation rates ($\hat\lambda_j = \lambda_{\max} = 3.0$) concentrate in the low-$N_{\pilot}$, short-$J_{\op}$ regime predicted by the admissibility analysis of Section~2.4 (Supplement~S4.1). The mean of the per-replication AnalyticSE distribution exceeds its median by two to three orders of magnitude for FFMLE Individual in the most fragile small-$N_{\pilot}$, short-$J_{\op}$ cells (e.g., mean $\approx 270$ versus median $0.58$ at the headline cell), indicating heavy upper tails in FFMLE's sampling distribution; the ratio drops to near unity by $N_{\pilot} = 100$, consistent with the finite-sample behavior reported by \citet{chenwang2016} (Supplement~S5.1). Truncation rates of the reported point estimates are larger for FFMLE Individual at small $N_{\pilot}$ than for the corrected estimators or Method A (Supplement~S4.2).

\section{Discussion}

Calibrating adaptively routed pilot items at operational scale forces a trade-off between full-information joint estimation---computationally burdensome and exposed to contamination from malfunctioning pilot items---and scalable point-estimate proxies that attenuate the recovered item parameters. Modular Local Calibration addresses this trade-off directly: the closed-form attenuation correction and Jacobian variance scaling derived in Section~2.4 reduce the framework's structural uncongeniality to a deterministic transformation of the local-regression estimator and its conditional first-order covariance. Across the simulation conditions, this transformation recovers approximately unbiased point estimates, achieves nominal-to-conservative coverage of 95\% confidence intervals throughout, and matches the OEM marginal-likelihood comparator on bias and MSE without the per-item numerical integration OEM requires.

\subsection{Methodological Contribution and Relation to Existing Methods}

The procedure brings three literatures into contact: Bayesian modularization, which partitions the joint model and blocks the feedback from pilot responses to the latent trait; multiple imputation via Plausible Values, which operationalizes that partition; and online item calibration, which supplies the substantive use case. The closed-form correction is the bridge between them, allowing Plausible-Values machinery to be applied to non-congenial local item calibration.

The trait estimate is fully blocked from the focal pilot's responses, a property shared with Method A and OEM but not with FFMLE-M-Method A, whose per-pilot Newton update incorporates focal-pilot responses into the score-equation evaluation, nor with MEM and joint FI MML, which fully restore the new-item-to-trait feedback channel. The contrast with FFMLE-M-Method A extends beyond this structural distinction to the assumptions each correction makes about the trait proxy. FFMLE assumes a classical measurement-error model whose zero-mean centering holds under random pilot routing but fails under the selective routing studied here, whereas the closed-form correction is constructed against the routed-subset posterior of $\theta_i$ and adapts to the marginal trait location and range-restriction effects that routing induces (Section~3.4.1).

The closed-form correction shares an asymptotic target with OEM's per-pilot quadrature but reaches it on the output side, so the trait-side computation (Plausible-Values generation, per-examinee operational posteriors) amortizes across the pilot pool as a byproduct of computations any operational scoring pipeline already performs. The Jacobian-scaled, Rubin-pooled variance construction (Proposition~\ref{prop:jacobian}) propagates operational item-parameter uncertainty into the new-item standard errors, a source of uncertainty that, to our knowledge, is treated as known and conditioned on in prior online-calibration frameworks. Finally, Firth's penalized likelihood applies natively to the local-regression form at the pilot-calibration stage (Section~3.4.4); analogous Jeffreys-type penalization is not routinely implemented in standard IRT software for joint FI MML, and OEM's integrated-marginal optimization does not natively accommodate it.

\subsection{Recommended Deployment}

The procedure is recommended for deployments in which pilot items are not used in operational scoring and routing depends only on operational responses. Under this configuration the imputation model conditions on operational responses alone, satisfying MAR for adaptive pilot routing, preventing tautological contamination, and preserving the modular structure that keeps per-pilot calibration tractable at scale. Within the family developed here, we recommend the Firth-penalized MLC-RR estimator as the default, on three grounds. First, MLC is preferred over PELC for inferential applications: PELC treats the EAP as an error-free predictor and produces residual undercoverage at moderate-to-large $N_j$, once local-regression variability no longer dominates the standard error (Section~3.4.1), whereas MLC propagates the operational posterior uncertainty through Rubin's between-imputation component. Second, among the three MLC pooling rules, RR is preferred for its principled finite-imputation variance penalty; RR and MN coincide on the point estimate and differ only in the variance inflation, and the stochastic EC rule tracks RR closely but requires larger $S$. Third, Firth's penalty reduces finite-sample bias and MSE at small $N_{\pilot}$---the sparse-routing setting in which separation is most likely---while remaining $O(1)$ relative to the log-likelihood and therefore asymptotically negligible and consistent with the population recovery of Proposition~\ref{prop:inverse} (Section~3.4.4). In combination, the closed-form correction, Jacobian variance scaling, and Firth's penalty produce approximately unbiased point estimates under the approximations of Section~2.3.1 (Propositions~\ref{prop:pelc_atten}--\ref{prop:inverse}) and nominal-to-conservative standard errors across the design grid studied.

\subsection{Limitations}

The methodology and simulation study have two main limitations with directions for further work. The analytic derivations and the simulation are restricted to the unidimensional 2PL. Extensions of these methods to alternate IRT models (such as the 3PL with estimated guessing, polytomous models, or multidimensional response models) are natural next steps. The correction structure (Gaussian convolution under the normal-ogive approximation, inverse mapping at empirical nuisance quantities) generalizes across models in principle under analogous approximations. The closed-form correction relies on a homoscedastic approximation to the posterior variance on the routed subset (Approximation~\ref{ass:homosc}); a heteroscedastic refinement conditioning on individual posterior variances could improve performance under severe within-subset heterogeneity.

The second limitation concerns the variance estimator. The analytic standard errors apply the delta method conditional on the realized routed subset and the empirical nuisance plug-ins (Proposition~\ref{prop:jacobian}). The unconditional sampling distribution adds contributions from variability in the plug-ins themselves and from their cross-covariance with the uncorrected pilot estimate, which is non-zero because both depend on the same operational Plausible Values \citep{mislevy1991}. The empirical higher-order analysis in Supplement~S6 decomposes the unconditional variance into three terms via a Taylor expansion of the corrected estimator with respect to the local-regression input and the nuisance plug-ins. Across the design grid, the cross-covariance contribution is uniformly negative and exceeds the (positive) nuisance-variance contribution in magnitude, leaving the analytic SE systematically conservative relative to the empirical SD by an asymptotically stable fraction of the variance rather than by a vanishing finite-sample effect. A sandwich-form refinement along the lines of \citet{murphytopel1985} or the two-step M-estimator framework of \citet{neweymcfadden1994} would propagate these contributions and is expected to recover an analytic SE that tracks the empirical SD at leading order. A refined variance estimator incorporating these terms is a clear direction for follow-up work.

\subsection{Extensions}

Two natural extensions reuse the framework's modular structure (operational calibration, Plausible-Values generation, local logistic regression) while varying the imputation model or post hoc treatment of the local estimates.

The first addresses deployments in which pilot responses themselves drive routing, as in continuous adaptive testing with feature-based predicted item parameters (e.g., the Duolingo English Test; \citealp{naismith2025, sharpnack2026}). Conditioning on operational responses alone then leaves selection into a pilot item's routed subset non-ignorable, because it omits the pilot responses that co-determined routing. Ignorability is restored by conditioning additionally on those responses, scored at the known feature-based parameters rather than the unobserved true ones. The practical obstacle is that a single Plausible-Values set cannot both exclude every pilot item, as contamination protection requires, and retain the pilot responses that routing depends on. A separate set per focal item resolves this tension by excluding only that item while conditioning on every item that routed to it, with the operational items scored at their calibrated parameters and the remaining pilots at their feature-based ones. An alternate approach that retains shared imputation is an empirical-Bayes imputation that includes all pilot items, with parameters drawn from the feature-based predictive distribution. Because the focal item then enters its own imputation, the uncongeniality dissolves, and with it the need for the analytic correction. In exchange, an empirical prior shrinks the focal estimate toward its feature-based prediction.

The second is the fully Bayesian errors-in-variables formulation noted in Section~2.3, which restores congeniality by treating individual traits as unobserved parameters within the local calibration, removing the need for the analytic correction at the cost of trait-side computation that does not amortize. Whether this formulation can be made computationally feasible at scale remains an open question; the empirical evidence reported here provides a benchmark against which such refinements can be measured.

\begin{singlespace}
\bibliographystyle{apacite}
\bibliography{references}
\end{singlespace}

\appendix
\section{Proofs of Propositions}
\label{app:proofs}

The proofs below rely on Approximations~\ref{ass:normal_ogive}--\ref{ass:joint_normal} (Section~2.3.1) and Condition~\ref{cond:adm} (Section~2.4) as labeled in the main text. Throughout, $\phi(\cdot;\,\mu, \sigma^2)$ denotes the density of $\mathcal{N}(\mu, \sigma^2)$, $\Phi(\cdot)$ denotes the standard normal CDF, and we use the Gaussian--probit convolution identity \citep{lord1968}
\begin{equation}
\int \Phi(A + Bx)\, \phi(x;\, 0, \sigma^2)\, dx \;=\; \Phi\!\left( \frac{A}{\sqrt{1 + B^2 \sigma^2}} \right),
\label{eq:gpconv}
\end{equation}
which can be verified by direct calculation.

\subsection*{Proof of Proposition~\ref{prop:pelc_atten} (PELC Attenuation)}

By the standard consistency theory for misspecified maximum likelihood (\citealp[][Theorem 2.5]{neweymcfadden1994}; \citealp{white1982}), the PELC estimator converges in probability to the value of $(a, d)$ that maximizes the population log-likelihood of the local logistic regression of $Y_{ij}$ on $\bar\theta_i$ over the routed subset. The first-order condition is satisfied at any $(a^*, d^*)$ for which $\mathrm{logit}^{-1}(a^* \bar\theta_i + d^*) = E[Y_{ij} \mid \bar\theta_i, i \in \mathcal{I}_j]$ on the support of $\bar\theta_i$. We identify this fixed point.

Under Approximations~\ref{ass:posterior_normal} and~\ref{ass:homosc}, $\theta_i \mid \bar\theta_i, i \in \mathcal{I}_j \stackrel{\cdot}{\sim} \mathcal{N}(\bar\theta_i, \sigma^2_{\post,j})$. Under Approximation~\ref{ass:normal_ogive},
\[
E[Y_{ij} \mid \bar\theta_i, i \in \mathcal{I}_j] \;\approx\; \int \Phi\!\left( \frac{a_j \bar\theta_i + d_j + a_j e}{D} \right) \phi(e;\, 0, \sigma^2_{\post,j})\, de.
\]
Applying \eqref{eq:gpconv} with $A = (a_j \bar\theta_i + d_j)/D$ and $B = a_j/D$ gives
\[
E[Y_{ij} \mid \bar\theta_i, i \in \mathcal{I}_j] \;\approx\; \Phi\!\left( \frac{(a_j \bar\theta_i + d_j)/D}{\sqrt{1 + (a_j/D)^2 \sigma^2_{\post,j}}} \right).
\]
Returning to the logistic metric via Approximation~\ref{ass:normal_ogive}, this equals $\mathrm{logit}^{-1}(a^* \bar\theta_i + d^*)$ with $a^* = a_j / \sqrt{1 + (a_j/D)^2 \sigma^2_{\post,j}}$ and $d^* = d_j / \sqrt{1 + (a_j/D)^2 \sigma^2_{\post,j}}$. The population first-order condition is therefore satisfied at $(a^*, d^*)$, establishing the claim. \hfill $\square$

\subsection*{Proof of Proposition~\ref{prop:mlc_atten} (MLC Attenuation)}

The argument parallels the PELC proof, with the conditional distribution adjusted to reflect that the predictor is now an imputed Plausible Value rather than the EAP. Under Approximation~\ref{ass:joint_normal}, $(\theta_i, \theta_i^{(s)}) \mid i \in \mathcal{I}_j$ is bivariate normal with marginal mean $\mu_{\sub,j}$, marginal variance $\sigma^2_{\sub,j}$, and correlation $\rho_j$. The conditional distribution is therefore
\[
\theta_i \mid \theta_i^{(s)}, i \in \mathcal{I}_j \;\stackrel{\cdot}{\sim}\; \mathcal{N}\!\left( \rho_j \theta_i^{(s)} + (1 - \rho_j)\mu_{\sub,j},\;\; \sigma^2_{\sub,j}(1 - \rho_j^2) \right).
\]
Decomposing $\theta_i = \rho_j \theta_i^{(s)} + (1-\rho_j)\mu_{\sub,j} + u_i$ with $u_i \mid \theta_i^{(s)} \sim \mathcal{N}(0, \sigma^2_{\sub,j}(1-\rho_j^2))$ and applying Approximation~\ref{ass:normal_ogive} yields
\[
E[Y_{ij} \mid \theta_i^{(s)}, i \in \mathcal{I}_j] \;\approx\; \int \Phi\!\left( \frac{a_j[\rho_j \theta_i^{(s)} + (1-\rho_j)\mu_{\sub,j}] + d_j + a_j u}{D} \right) \phi(u;\, 0, \sigma^2_{\sub,j}(1-\rho_j^2))\, du.
\]
Applying \eqref{eq:gpconv} with $A = (a_j[\rho_j \theta_i^{(s)} + (1-\rho_j)\mu_{\sub,j}] + d_j)/D$ and $B = a_j/D$, and converting the result back to the logistic metric via Approximation~\ref{ass:normal_ogive}, gives a logistic predictor with slope $a_j \rho_j / \sqrt{1 + (a_j/D)^2 \sigma^2_{\sub,j}(1-\rho_j^2)}$ and intercept $(d_j + a_j(1-\rho_j)\mu_{\sub,j}) / \sqrt{1 + (a_j/D)^2 \sigma^2_{\sub,j}(1-\rho_j^2)}$. The same misspecified-MLE consistency argument identifies these as the population limits of the MLC estimator, applying identically to all three pooled MLC estimators (RR, MN, EC), each of which converges in probability to the same population limit determined by the local-regression first-order condition at the population marginal response function. \hfill $\square$

\subsection*{Proof of Proposition~\ref{prop:inverse} (Inverse Mapping)}

We verify the four identities directly. The derivation requires only $a_j \ne 0$; the $a_j = 0$ case follows by continuity, with the corrections reducing to the identity. The corrections are sign-equivariant: the population limit $a_\dagger$ inherits the sign of $a_j$ (since the attenuating denominator is always positive), the expansion factors $\lambda_{\PELC,j}$ and $\lambda_j$ are positive functions of $a_j^2$, and the corrected estimators $\lambda \cdot \hat a$ inherit the sign of the local-regression coefficient. The 2PL setting conventionally takes $a_j > 0$, but no step of the proof depends on the sign.

\textit{PELC.} Let $(a_\dagger, d_\dagger)$ denote the population limits of the PELC estimators identified in Proposition~\ref{prop:pelc_atten}. From $a_\dagger^2 = a_j^2 / (1 + (a_j/D)^2 \sigma^2_{\post,j})$,
\[
1 - a_\dagger^2\, \sigma^2_{\post,j} / D^2 \;=\; \frac{1 + (a_j/D)^2 \sigma^2_{\post,j} - (a_j/D)^2 \sigma^2_{\post,j}}{1 + (a_j/D)^2 \sigma^2_{\post,j}} \;=\; \frac{1}{1 + (a_j/D)^2 \sigma^2_{\post,j}}.
\]
Therefore $\lambda_{\PELC,j}^{\,2} = 1 + (a_j/D)^2 \sigma^2_{\post,j} = a_j^2 / a_\dagger^2$, so $\lambda_{\PELC,j} a_\dagger = a_j$. The intercept identity follows because $d_\dagger / a_\dagger = d_j / a_j$ from Proposition~\ref{prop:pelc_atten}, so $\lambda_{\PELC,j} d_\dagger = (d_j / a_j) \lambda_{\PELC,j} a_\dagger = d_j$. Condition~\ref{cond:adm} ensures $\lambda_{\PELC,j}$ is real and finite.

\textit{MLC.} Let $(a_\dagger, d_\dagger)$ denote the population limits of the MLC estimators from Proposition~\ref{prop:mlc_atten}, and let $c_j = \sigma^2_{\sub,j}(1-\rho_j^2)/D^2$. From $a_\dagger^2 (1 + a_j^2 c_j) = a_j^2 \rho_j^2$, multiplying out gives
\[
a_j^2 (\rho_j^2 - c_j a_\dagger^2) \;=\; a_j^2 \rho_j^2 - a_j^2 c_j a_\dagger^2 \;=\; a_\dagger^2(1 + a_j^2 c_j) - a_\dagger^2 a_j^2 c_j \;=\; a_\dagger^2.
\]
Therefore $\lambda_j^{\,2} = (\rho_j^2 - c_j a_\dagger^2)^{-1} = a_j^2 / a_\dagger^2$, so $\lambda_j a_\dagger = a_j$. Equivalently, $\sqrt{1 + a_j^2 c_j} = \lambda_j \rho_j$. From Proposition~\ref{prop:mlc_atten}, $d_\dagger = (d_j + a_j(1-\rho_j)\mu_{\sub,j})/(\lambda_j \rho_j)$. Substituting into the MLC intercept correction,
\[
\rho_j \lambda_j d_\dagger - \lambda_j(1-\rho_j)\mu_{\sub,j} a_\dagger \;=\; (d_j + a_j(1-\rho_j)\mu_{\sub,j}) - (1-\rho_j)\mu_{\sub,j} a_j \;=\; d_j,
\]
where we used $\lambda_j a_\dagger = a_j$.

Empirical convergence of both the PELC and MLC corrections to $(a_j, d_j)$ in probability follows by the continuous mapping theorem, since the plug-ins converge to their population analogues in the asymptotic setting of Section~2.5 and Condition~\ref{cond:adm} holds in the limit, placing the population radicands strictly inside the domain on which the correction map is continuous. \hfill $\square$

\subsection*{Proof of Proposition~\ref{prop:jacobian} (Conditional Smooth-Branch Delta-Method Covariance)}

\textit{Jacobian.} Differentiate $\hat\lambda_j^{-2} = \hat\rho_j^2 - \hat c_j \hat a_{\MLC}^2$ with respect to $\hat a_{\MLC}$, holding the nuisance plug-ins fixed: $-2 \hat\lambda_j^{-3} \, (\partial \hat\lambda_j / \partial \hat a_{\MLC}) = -2 \hat c_j \hat a_{\MLC}$, so $\partial \hat\lambda_j / \partial \hat a_{\MLC} = \hat c_j \hat a_{\MLC} \hat\lambda_j^{\,3}$. By the product rule,
\begin{align*}
J_{11}^{\text{smooth}} &= \frac{\partial(\hat\lambda_j \hat a_{\MLC})}{\partial \hat a_{\MLC}} = \hat c_j \hat a_{\MLC}^2 \hat\lambda_j^{\,3} + \hat\lambda_j = \hat\lambda_j(1 + \hat c_j \hat a_{\MLC}^2 \hat\lambda_j^{\,2}).
\end{align*}
From $\hat\lambda_j^{\,2}(\hat\rho_j^2 - \hat c_j \hat a_{\MLC}^2) = 1$, we have $\hat c_j \hat a_{\MLC}^2 \hat\lambda_j^{\,2} = \hat\rho_j^2 \hat\lambda_j^{\,2} - 1$, so $1 + \hat c_j \hat a_{\MLC}^2 \hat\lambda_j^{\,2} = \hat\rho_j^2 \hat\lambda_j^{\,2}$, yielding $J_{11}^{\text{smooth}} = \hat\rho_j^2 \hat\lambda_j^{\,3}$. The remaining derivatives follow directly: $J_{12}^{\text{smooth}} = 0$ since neither $\hat\lambda_j$ nor $\hat a_{\MLC}$ depends on $\hat d_{\MLC}$; $J_{22}^{\text{smooth}} = \hat\rho_j \hat\lambda_j$ by inspection of \eqref{eq:dcorr_mlc}; and
\[
J_{21}^{\text{smooth}} = \hat\rho_j \hat d_{\MLC} (\hat c_j \hat a_{\MLC} \hat\lambda_j^{\,3}) - (1-\hat\rho_j)\hat\mu_{\sub,j} \cdot J_{11}^{\text{smooth}} = \hat\lambda_j^{\,3} \hat\rho_j \!\left( \hat d_{\MLC} \hat a_{\MLC} \hat c_j - \hat\mu_{\sub,j} \hat\rho_j(1-\hat\rho_j) \right).
\]
The PELC formulas follow analogously by differentiating $\hat\lambda_{\PELC,j}^{\,-2} = 1 - \hat a_{\PELC}^2 \hat\sigma^2_{\post,j}/D^2$.

\textit{Conditional covariance interpretation.} Write $\hat\eta_{j,\corr} = g(\hat\eta_j, \hat\nu)$, where $\hat\eta_j$ is the local-regression estimator and $\hat\nu = (\hat\rho_j, \hat\mu_{\sub,j}, \hat\sigma^2_{\sub,j}, \hat\sigma^2_{\post,j})$. Conditional on $\hat\nu$ and the realized subset, $\hat\eta_j$ has covariance $\mathbf{V}_{\uncorr}$, and a first-order Taylor expansion of $g(\cdot, \hat\nu)$ around $\eta_j^* = \mathrm{plim}(\hat\eta_j \mid \hat\nu)$ gives the conditional covariance $\mathbf{J}\, \mathbf{V}_{\uncorr}\, \mathbf{J}^\intercal$. The unconditional distribution additionally involves variation in $\hat\nu$:
\[
\sqrt{N_j}\,(\hat\eta_{j,\corr} - \eta_j) \;=\; \mathbf{J}\, \sqrt{N_j}\,(\hat\eta_j - \eta_j^*) \;+\; \frac{\partial g}{\partial \nu^\intercal}\, \sqrt{N_j}\,(\hat\nu - \nu) \;+\; o_p(1).
\]
The routed-subset moments are computed from the same $N_j$ examinees as the local regression, so $\sqrt{N_j}(\hat\nu - \nu) = O_p(1)$ and the nuisance term contributes at leading order. The conditional covariance therefore omits leading-order contributions from nuisance variability and the cross-covariance between $\hat\eta_j$ and $\hat\nu$ (non-zero by construction since both depend on the routed-subset Plausible Values). A sandwich-form variance accommodating these contributions is the standard remedy \citep{murphytopel1985, neweymcfadden1994}; the unconditional behavior, including cross-term sign, is examined empirically in Section~3.4.5. \hfill $\square$

\end{document}